\documentclass[prl,aps,twocolumn,showpacs,preprintnumbers,amsmath,amssymb,floatfix,superscriptaddress]{revtex4}

\usepackage{natbib}
\usepackage{graphicx}
\usepackage{dcolumn}
\usepackage{bm}
\usepackage{textcomp}
\usepackage{units}
\usepackage{longtable}
\usepackage{color}
\usepackage{amsmath}

\begin{document}


\title{An experimental and theoretical guide to strongly interacting Rydberg gases}

\author{Robert L\"ow}
\email{r.loew@physik.uni-stuttgart.de}
\affiliation{5. Physikalisches Institut, Universit\"at Stuttgart, 70569 Stuttgart, Germany }
\author{Hendrik Weimer}
\affiliation{Physics Department, Harvard University, 17 Oxford Street, Cambridge, MA 02138, USA} %
\affiliation{ITAMP, Harvard-Smithsonian Center for Astrophysics, 60 Garden Street, Cambridge, MA 02138, USA}%
\author{Johannes Nipper}
\affiliation{5. Physikalisches Institut, Universit\"at Stuttgart, 70569 Stuttgart, Germany }
\author{Jonathan B. Balewski}
\affiliation{5. Physikalisches Institut, Universit\"at Stuttgart, 70569 Stuttgart, Germany }
\author{Bj\"orn Butscher}
\affiliation{5. Physikalisches Institut, Universit\"at Stuttgart, 70569 Stuttgart, Germany }
\author{Hans Peter B\"uchler}
\affiliation{Institut f\"ur Theoretische Physik III, Universit\"at Stuttgart, 70569 Stuttgart, Germany  }
\author{Tilman Pfau}
\affiliation{5. Physikalisches Institut, Universit\"at Stuttgart, 70569 Stuttgart, Germany }

\date{\today}

\begin{abstract}
 We review experimental and theoretical tools to excite, study and understand strongly interacting Rydberg gases. The focus lies on the excitation of dense ultracold atomic samples close to, or within quantum degeneracy, to high lying Rydberg states. The major part is dedicated to highly excited S-states of Rubidium, which feature an isotropic van-der-Waals potential. Nevertheless are the setup and the methods
 presented also applicable to other atomic species used in the field of laser cooling and atom trapping.

\end{abstract}

\maketitle

\section{\label{chapintroduction} Introduction}

The field of Rydberg atoms has been revolutionized by
the combination with ultracold atomic gases. The availability of commercial high power diode laser systems allow for efficient excitation into electronically highly excited states with excellent frequency resolution. Here the major motivation behind many experiments is to make use of the strong interaction mechanism among Rydberg atoms, which become especially apparent at large atomic densities as found in ultracold gases. On the other hand the precise spectroscopy of the interaction energies is supported by the availability of trapped atoms, the almost negligible Doppler effect given by small temperatures and the possibility to prepare all atoms in a specific quantum state with high fidelity. Combining this control over internal and external degrees of freedom for atoms, molecules and ions with switchable long range interactions establishes an ideal basis for applications in quantum simulation, quantum computation and quantum optics.

Generally there exist two pathways to prepare a controllable ensemble
of strongly interacting particles. One way is to start with a system
of weakly interacting particles as in the case of quantum degenerate
gases and then try to implement strong interactions among them
e.g. with Feshbach resonances \cite{Zwierlein:2005}, optical lattices
\cite{Greiner:2002}, high finesse resonators \cite{Baumann:2010}, or
by extreme cooling in order to make the initially weak interaction
energy the dominant scale \cite{Jo:2009}. The second approach utilizes
already strongly interacting particles as polar molecules
\cite{Ni:2008} and ions, where the latter already constitutes the best
working quantum computer \cite{Monz2011,Home2009} and
universal quantum simulator \cite{Lanyon2011} in the
world. The drawback of the latter approach is, that
the presence of interactions requires the particles
  to be restricted to specific geometries.

At this point Rydberg states might be an ideal add-on to both approaches, since they are switchable and exhibit interaction strengths comparable to ionic systems \cite{Saffman:2010}. \\
The major interaction mechanisms between two highly excited atoms at large distances is typically given by the van-der-Waals interaction and for smaller distances by the dipole-dipole interaction \cite{Saffman:2010}. The effects of both interaction types have been observed in laser cooled clouds \cite{Afrousheh:2006,Anderson:1998,Carroll:2006,Liebisch:2005,Tong:2004,Singer:2004,Vogt:2007,Ditzhuijzen:2008,Pritchard:2010} and also in optically and magnetically trapped clouds \cite{Heidemann:2007,Nipper:2011,Cooper:2009}. A prominent effects of the strong interaction mechanism is an excitation blockade in which the energy of a Rydberg state is shifted out of resonance with respect to the exciting laser by a neighboring Rydberg atom.

The most drastic consequence of the dipole blockade
is the occurance of a quantum phase transition to a crystalline phase of Rydberg excitations
\cite{Weimer:2008,Weimer:2010b,Sela:2011}. The parameters controlling
the transition can be tuned with external laser fields. While this
Rydberg crystal has so far eluded direct observation, critical
fluctuations of this quantum phase transition have already been
reported \cite{Loew:2009} in an non adiabatic manner. The realization of the crystalline phase, which is the ground state of the system, would require adiabaticity during preparation \cite{Pohl:2010,Bijnen:2011,Schachenmayer:2010,Weimer:2011b}.

The blockade effect is also used in a more involved approach with exactly two atoms located in a certain distance establishing the prototype of a quantum gate \cite{Urban:2009,Gaetan:2009}. Since the loading of individual sites is still probabilistic \cite{Schlosser:2001}, the scalability of this approach is limited as long as there is no deterministic loading scheme available \cite{Saffman:2002,Kuhr:2001}. An alternative way to achieve a large quantum register with individually resolved sites is a Mott-insulator state with single site resolution \cite{Sherson:2010,Bakr:2009} or a switchable local electric field, e.g. produced by an electron beam \cite{Gericke:2008} which allows an addressing via the spatial dependent Stark shift. \\
Although the first proposal to use the interaction between two spatially resolved Rydberg atoms \cite{Jaksch:2000} seemed quite involved in the year 2000, it triggered a large theoretical effort and several experimental groups started to implement these ideas. There exist elaborate proposals for the realization of a universal quantum simulator \cite{Weimer:2010,Weimer:2011,Brion:2007} and various ways of to implement a quantum computer \cite{Jaksch:2000,Lukin:2001,Moller:2008,Saffman:2008,Saffman:2010}. Some of these schemes rely on collective states, where one Rydberg excitation is shared by several atoms \cite{Lukin:2001,Muller:2009,Brion:1999}. In this article we will focus on the many particle character of the collective states in an ensemble of ultracold atoms with extensions well larger than the blockade radius. A detailed understanding of the excitation dynamics in these systems  into a strongly correlated quantum system, its microscopic description, the limitations of the effective models used, the sources of decoherence and dephasing and finally of the applicable experimental techniques will not only be helpful for the implementation of various quantum simulation and quantum computation protocols but also for other tasks involving interacting Rydberg atoms. One could be the idea of combining interacting Rydberg atoms with quantum degenerate gases \cite{Honer:2010,Pupillo:2010,Henkel:2010,Maucher:2011} to render new interaction mechanisms. Besides the effects of the strong interaction on the material wavefunctions there exists also a strong backaction on the radiation field, which can produce and also analyze various kinds of non classical light fields \cite{Honer:2011,Pohl:2010,Pederson:2009}.  \\
In the following we will first describe the basic properties of non-interacting Rydberg atoms and deduce from this the applicable energy and time scales for meaningful experiments. These considerations also set the preconditions and limitations to the techniques for the excitation of ultracold trapped atoms to Rydberg states and their subsequent detection. In the next section we will include the interaction among Rydberg states and will derive a meanfield model for the strongly interacting case, which describes quite well the experimental findings. Finally we will summarize experimental tools and results on strongly interacting quantum systems based on Rydberg atoms.

\section{\label{ChapRydGeneral} General properties of Rydberg atoms}

The properties of Rydberg atoms are similar to that of ground state atoms, but are severely enlarged with the principal quantum number $n$ \cite{Gallagher:1994}. Table \ref{RydProps} shows an overview of some important values and their corresponding scaling.

\begin{table*}[htb]
\begin{center}
\begin{tabular}{|l|l|l|l|l|}
\hline
Property & Expression & $(n^\star)^x$ & Rb(5S) - ground state & Rb(43S) - Rydberg state\\
\hline
Binding energy & $E_{n^\star}=-\frac{Rhc}{(n^\star)^2}$ & $(n^\star)^{-2}$ & \unit[4.18]{eV} & \unit[8.56]{meV} \\
\hline
Level spacing & $E_{n^\star}-E_{n^\star+1}$ & $(n^\star)^{-3}$ & \unit[2.50]{eV} (5S-6S)& \unit[413.76]{$\mu$eV} = \unit[100.05]{GHz} (43S-44S)\\
\hline
Orbit radius & $ \langle r \rangle=\frac{1}{2} \left( 3(n^\star)^{2}-l(l+1) \right)$ & $(n^\star)^{2}$ & \unit[5.632]{a$_0$} \cite{Radzig:1985} & \unit[2384.2]{a$_0$} \\
\hline
Polarizability & $ \alpha $ & $(n^\star)^{7}$ & \unit[-79.4]{mHz/(V/cm)$^2$ \cite{CRC:2004}} & \unit[-17.7]{MHz/(V/cm)$^2$} \\
\hline
Lifetime (spont. decay) & $ \tau=\tau\prime\cdot(n^\star)^{\gamma} $ & $(n^\star)^{3}$ & 5P$_{3/2}$-5S$_{1/2}$: 26.2ns  & \unit[42.3]{$\mu$s} at 300K incl. BBR\\
\hline
transition dip. moment & $ \langle 5P\| er\| nS \rangle $ & $(n^\star)^{-1.5}$ & 5S$_{1/2}$-5P$_{3/2}$: 4.227 ea$_0$ & 5P$_{3/2}$-43S${_1/2}$: 0.0103 ea$_0$ \\
\hline
transition dip. moment & $ \langle nP\| er \| (n+1)S \rangle $ & $(n^\star)^{-2}$ & --- & 43P$_{3/2}$-43S$_{1/2}$: 1069 ea$_0$  \\
\hline
van der Waals coeff. & $ C_6 $ & $(n^\star)^{11}$ & 4707 a.u. \cite{Claussen:2003} & -1.697 10$^{19}$ a.u. \\
\hline
\end{tabular}
\caption{\label{RydProps} With some explicit values for the 43S state in Rubidium it is possible to get an idea for the framework and the boundary conditions for an experimental setting involving Rydberg atoms. With the given scaling laws $(n^\star)^x$ one can also find an estimate for the physical situation with other Rydberg states.}
\end{center}
\vspace{-0.6cm}
\end{table*}

\subsection{\label{RbRydberg} Electronic structure and lifetimes of highly excited Rubidium atoms}

The largest fraction of all laboratories working with ultracold atoms use Alkalis, since their simple hydrogen-like structure makes their excited states accessible and controllable with only a few laser frequencies. For most of them the deviation from a hydrogen atom is caused by the core electrons which add a repulsive Coulomb potential for the Rydberg electron and a simple $1/r$ Coloumb potential is not applicable anymore. This effect can be faced in a phenomenological manner by introducing a quantum defect $\delta(n,j,l)$, which results in an altered Rydberg formula

\begin{equation}\label{eq.quantdefect}
E(n,j,l)=-\frac{R^\prime}{(n-\delta(n,j,l)^2)}=-\frac{R^\prime}{(n^\star)^2},
\end{equation}

where $R^\prime$ is the specific Rydberg constant for the element in question, e.g. $R^\prime=109736.605$ cm$^{-1}$ for Rubidium ($R^\prime=109737.316$ cm$^{-1}$ for hydrogen). The effective quantum number $n^\star$ mostly depends on $l$ and is in lowest order for Rubidium determined by $\delta(l=0)=3.13,\,\delta(l=1)=2.64,\,\delta(l=2)=1.35,\,\delta(l=3)=0.016\,$ and $\delta(l>3)\approx 0$ \cite{Li:2003,Jianing:2006,Afrousheh:2006b}. More accurate values for nS and nD states have been recently measured \cite{Mack:2011}. As a consequence of the l-dependent quantum defects the degeneracy of the s,p,d and f-states is lifted and only states with $l>3$ show a hydrogenlike behavior. With the knowledge of the energies of the individual excited states it is possible to solve the Schr\"odinger equation analytically \cite{Kostelecky:1985} or numerically to obtain also the radial wavefunctions. For that purpose one has to add a term $V_p=-\frac{\alpha_d}{2r^4}$ to account for the Coulomb potential of the core electrons in terms of a core polarizability e.g. $\alpha_d=9.023$ a.u. in the case of Rubdium \cite{Litzen:1970}.
Since the probability of finding a Rydberg electron inside the core is quite small the quality of calculated wavefunctions are often of sufficient accuracy to be used to reliably compute dipole matrix elements, while improved model potentials exist for other cases \cite{Marinescu:1994}. These dipole matrix elements are crucial for the calculation of the excited state lifetime, the polarizability, the van-der-Waals interaction and many more properties \cite{Gallagher:1994}. \\

\begin{figure}
\includegraphics[width=\linewidth]{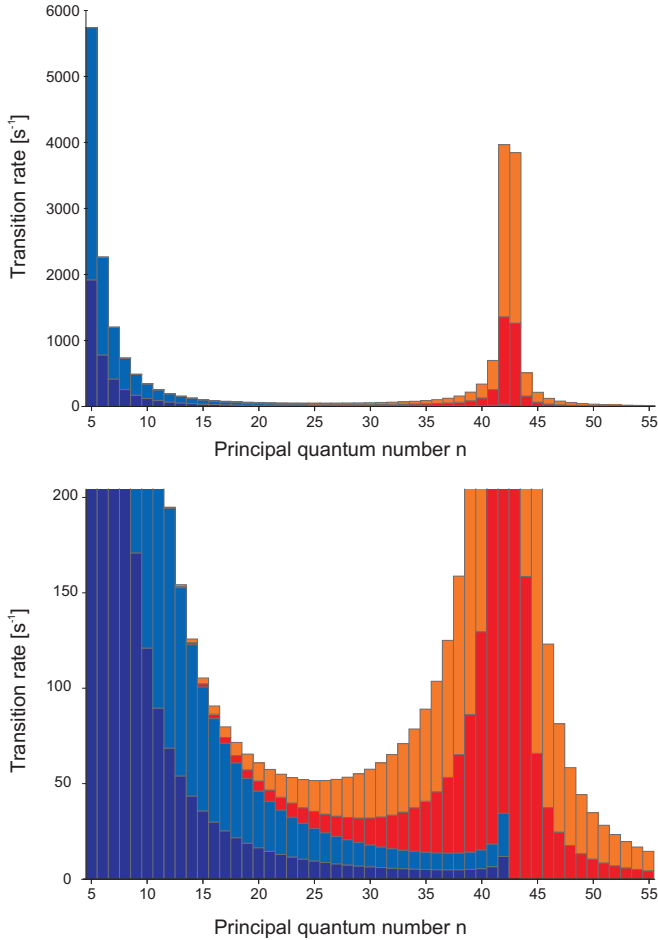}
\caption{\label{pic_EinsteinA} Stacked bar chart for the calculated transition rates of the $43S_{1/2}$ state of Rubidium (top) and a 30x enlarged view (bottom). The Einstein A coefficients for the spontaneous decay are shown in blue $nP_{1/2}$ and light blue $nP_{3/2}$ to all energetically lower lying states $(5\leq n\leq 42)$. The sum of these coefficients results in a lifetime due to spontaneous decay of 80.8 $\mu s$. An additional loss of population is caused by blackbody radiation which favors neighboring states shown in red $nP_{1/2}$ and orange $nP_{3/2}$. The total lifetime (including blackbody radiation at 300K) is then reduced to 42.3 $\mu s$, which corresponds to a Lorentzian linewidth of 3.76 kHz and  is in good agreement with more elaborate calculations carried out in \cite{Beterov:2009}.}
\end{figure}

The lifetime of Rydberg states is determined by the radiative decay to lower lying levels but also by transitions to higher (and lower) lying states induced by blackbody radiation as in shown in Fig. \ref{pic_EinsteinA}. The major contribution to the lifetime is given by spontaneous decay events to low lying levels where the product of interaction strength and density of states of the electromagnetic field scaling as $\omega^3$ outweighs the decreasing dipole moments ($\sim n^{\star {-1.50}}..n^{\star {-1.55}}$). For Rydberg state $n\gtrsim 40$ the absorption and stimulated emission of thermally occupied infrared modes to neighboring states starts to dominate the lifetime. The latter effect can be eliminated by using a cryogenic environment as it is done in cavity QED experiments with Rydberg atoms \cite{Raimond:2001}.\\
The transition rate between two states $i$ and $f$ with a dipole matrix element $\langle i|r|f\rangle$, separated by the energy $\hbar \omega_{if}$, is typically expressed in terms of an Einstein $A$ coefficient for the spontaneous decay

\begin{equation}\label{eq.einsteinA}
A=\frac{2 e^2 \omega_{if}^3}{3 \epsilon_0 c^3 h} |\langle i|r|f\rangle|
\end{equation}

and an Einstein $B$ coefficient ($B=A N(\omega)$) for stimulated emission as well as absorption of black body radiation.
The number of black body photons $N(\omega)$ per mode at a given temperature $T$ is given by

\begin{equation}\label{eq.bbr}
N(\omega)=\frac{1}{e^{\hbar \omega/k_B T}-1},
\end{equation}

which results in a total lifetime $\tau^{-1}=\sum A+\sum A N(\omega)$. The first sum is limited to states with energies below the Rydberg state and the second sum runs over all allowed dipole transitions, however neglecting transitions to ionized states. The actual measurement of the radiative lifetime by delayed field ionization and detection of remaining Rydberg states is quite involved since one has to distinguish the Rydberg state in question from energetically close by states, which are likely to be populated by blackbody radiation \cite{Branden:2010}. Here this problem has been avoided by a state sensitive detection scheme. Another method employs coherent Ramsey spectroscopy, which is only sensitive to one specific Rydberg state \cite{Butscher:2010} and also delivers reliable lifetimes if the total linewidth of the excitation lasers are accordingly small.

\subsection{\label{ChapEFields} Rubidium Rydberg atoms in electric fields}

\begin{figure}
\includegraphics[width=\linewidth]{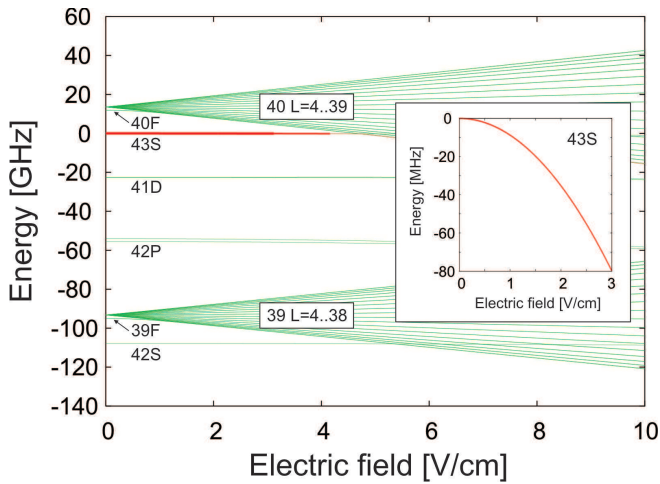}
\caption{\label{pic_stark43s} Calculated Stark map including LS coupling with respect to the $43S_{1/2}$ state in $^{87}$Rb. Only states with $L\geq 4$ exhibit hydrogen like linear Stark fans. For lower $L$ the states are shifted quadratically as it can be seen in the inset for the $43S_{1/2}$ state.}
\end{figure}

The large polarizability of Rydberg states is a welcome feature for experimentalists, since it allows for simple and fast manipulation of the excited state energies by applying small electric fields $E$. With the knowledge of the dipole matrix elements it is straightforward to find the new eigenenergies in the presence of an electric field \cite{Zimmerman:1979}. The corresponding Stark map, as shown in Fig. \ref{pic_stark43s}, exhibits two kinds of Stark shifted states. If the degeneracy is lifted by the quantum defect, the Stark effect starts of quadratic $E_{Stark}=\alpha E^2/2$ with the polarizabilty $\alpha$, and is linear in the case of degeneracy, which corresponds to a permanent dipole moment. In the case of Rubidium the polarizability for the 43S states is $\alpha=-17.7$ MHz/(V/cm)$^2$, which shifts this state at a field of only 1 V/cm already 350 linewidths to the red. For states with higher angular momentum the manifold of magnetic substates discloses a unique pattern in the presence of an electric field \cite{Grabowski:2006}, which can be used to calibrate the actual electric field \cite{Kuebler:2010}.

\subsection{\label{ChapWW} Interaction of Rydberg atoms}

In the case of large distances $r$ between two Rydberg atoms the interaction is in most cases determined by a van der Waals interaction $V_{vdW}=C_6/r^6$. The van der Waals interaction is theoretically best described by a second order effect in perturbation theory of the dipole-dipole interaction ($\sim r^{-3}$) among two atoms, which leads to a scaling $\sim r^{-6}$. In zero electric field any atom in a single parity eigenstate does not exhibit a permanent dipole moment. Nevertheless, fluctuations of the electron distribution lead to momentary dipole moments $\sim r^3$, which induce dipole moments in the other atom. In second order perturbation theory the interaction energy is proportional to the square of the dipole moment for each atom divided by the energy spacing of the pair states \cite{Reinhard:2007}. The largest contribution is given by the dipole moments to close by Rydberg states, which results in a scaling behavior with respect to the principal quantum number as $C_6 \sim n^{11}$. A numerical evaluation of the interaction potentials gives e.g. for Rubidium for two Rydberg atoms in a S-state and $30<n<95$ \cite{Singer:2005} in atomic units a simple scaling law

\begin{equation}\label{eq.c6singer}
C_6=n^{11}\left( 11.97-0.8486 n + 3.385\times10^{-3} n^2 \right).
\end{equation}

 Various interaction parameters have been calculated with pertubative methods \cite{Reinhard:2007} but also by diagonalization \cite{Schwettman:2007,Cabral:2011,Samboy:2011}. For Rubidium in the 43S state the van der Waals coefficient is $-1.697\times 10^{19}$ a.u. which is not only much larger than in the ground state 5S ($C_6=+4707$ a.u.) but is also opposite in sign resulting in a repulsive potential. This is a quite useful feature since attractive forces eventually lead to inelastic collisions and ionization \cite{Walz:2004,Robinson:2000,Li:2005,Pohl:2003,Zhang:2008}. At small distances the van der Waals interaction becomes comparable to the energy differences between the relevant pair states, the F\"orster defect, and pertubative methods are not valid anymore. At smaller distances the interaction potential adopts a dipolar $\sim 1/r^3$ character, which is valid down to the LeRoy radius where the exchange interaction of the electrons cannot be neglected anymore. For the 43S state in Rubidium the F\"orster defect is 3 GHz \cite{Walker:2008} which corresponds to the van-der-Waals interaction energy at a distance just below one micron. However, the distance between two Rubidium atoms in evaporatively cooled clouds may easily reach smaller values leading to interaction energies of many tens of GHz, an energy much larger than all other relevant energies scales in the system. Even at a distance of 5 $\mu$m the interaction energy is still 150 kHz, which establishes in combination with a comparable coupling strength of the driving laser fields and sufficiently dense atomic samples the domain of strongly interacting Rydberg gases. \\

\section{\label{ChapExpBasics} Rydberg excitation of ultracold atomic gases}

For experiments with Rydberg atoms in ultracold gases additional equipment and methods for excitation and detection of Rydberg states are required, without affecting the cooling and trapping procedure. The required level of performance of these new tools is determined by the properties of the Rydberg atoms. For example the laser light for the Rydberg excitation has to be sufficiently intense and narrow in frequency to achieve a coherent evolution within the natural lifetime. The 43S in Rubidium has a lifetime (including BBR at 300K) of 42.3 $\mu$s which corresponds to a linewidth of 3.76 kHz and the excitation laser should be accordingly narrow. The lifetime also sets the timescale for the experiments and by this the desired coupling strength of the laser system. To observe e.g. a full Rabi-oscillation within the radiative lifetime of the 43S state in Rubidium, the Rabi frequency $\Omega_R$ has to be accordingly strong and the linewidth of the excitation sufficiently narrow. Given that the dipole matrix elements for excitation into Rydberg states are typically quite small, a sufficient strong laser source with several tens of milliwatts is appropriate. But many proposals e.g. for quantum computing and quantum simulation will only be manageable with several hundred milliwatts of available laser power. \\
After excitation one also wants to detect the produced Rydberg atoms efficiently. Absorptive and fluorescing methods, as typically used for ground state atoms, are impractical since the spontaneous scattering rates are roughly a thousand times smaller. One exception are Rydberg states of alkaline earths, where the second non-excited electron can be used for optical detection \cite{Simien:2004}. The most common method relies on efficient and fast ion detection with mutlichannel plates or channeltrons, which can reach counting efficiencies of 90\%. To do so one has to field ionize the Rydberg states first either with a strong electric field or by a far infrared laser pulse \cite{Potvliege:2004}. The field plates inside the vacuum chamber to ionize the atoms and to guide them to the ion detectors are also very useful to apply moderate electric fields to shift the Rydberg levels in and out of resonance with the driving laser fields, to subtract unwanted ions or to implement a Ramsey interferometer \cite{Ryabtsev:2003}. \\
The other contributor, ultracold atomic gases, are 15 years after the fist observation of Bose-Einstein condensation \cite{Ketterle:2002,Cornell:2002} of the shelf in many atomic physics laboratories. There exist countless tools and methods to control the internal and external degrees of freedom, where most of them also get along with the technical constraints of highly excited Rydberg states. For precision spectroscopy trapped gases at extremely low temperatures are an ideal sample. The Doppler-effect is largely reduced, the interaction time with the probing light field is limited either by the excited state lifetime or by the lifetime of the trap, which can be minutes. Actually on the typical timescales of Rydberg experiments of a few microseconds the ultracold atoms move only a fraction of the wavelength of the exciting lasers, which has led to the expression of frozen Rydberg gases. Additionally to this the high densities of fully polarized samples make the interaction potential among Rydberg states accessible to a spectroscopic precision limited only by the excited state lifetime. Such measurements can also be taken in zero electric and magnetic fields by switching of all trapping potentials (magnetic or dipole) during excitation.

\subsection{\label{ChapTwoPhoton} Two-photon excitation of ultracold atoms into Rydberg states}

\begin{figure}
\includegraphics[width=\linewidth]{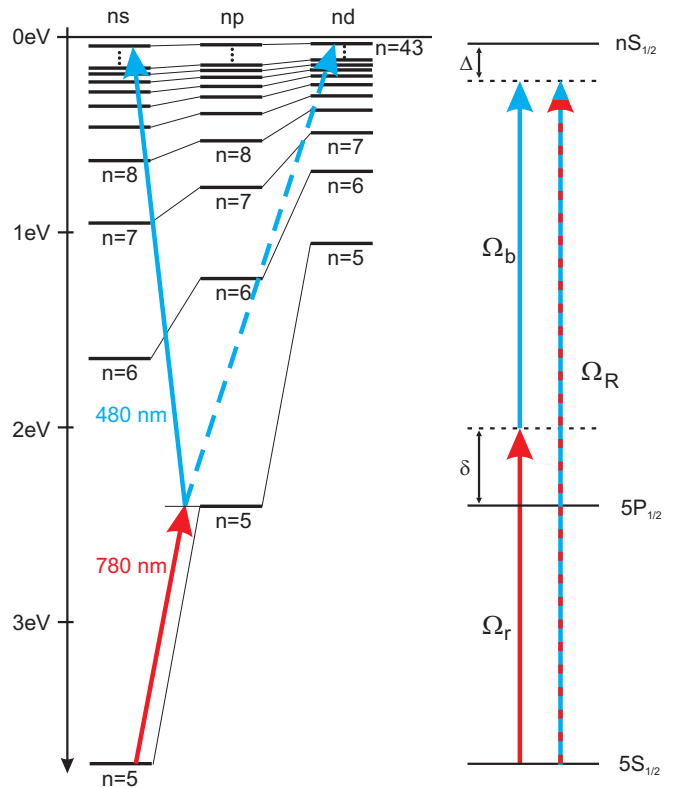}
\caption{\label{exc} The left part of the picture depicts the level scheme of Rubidium, which is well described by the quantum defects as described in the text. Due to angular momentum conservation it is only possible to reach s- and d-states with two photons when starting from the 5s ground state. In all experiments we excite the atoms from the 5s F=2,mF=2 via the 5p F=3 mf=3 to the 43s j=1/2 state. Although the hyperfine splitting of the 43s state is roughly 500 kHz \cite{Li:2003,Mack:2011} we do not observed the $F^\prime=1$ line due to the selection rules of the given excitation scheme. This three-level system can now be reduced to an effective two level system by introducing a large detuning $\delta$, which largely simplifies the subsequent modeling.}
\end{figure}

The excitation into Rydberg states with principal quantum numbers
ranging from $n=20$ up to the ionization threshold is usually accomplished by
a two-photon- and sometimes by three-photon excitation schemes. For a single-photon excitation of Rubidium to Rydberg states one
would have to use light at 297 nm, which is difficult to produce and would also not allow to access the desired S-states.
In Rubidium there are in principle two two-photon excitation schemes feasible, here we use light at 780 nm and 480 nm for to the 5s-5p-ns/nd scheme as shown in Fig. \ref{exc} but also the alternative route 5s-6p-ns/nd with 420 nm and 1016 nm lies in the range of standard diode laser systems.

In all experiments, we detune the red laser far from resonance to minimize spontaneous scattering of photons via population of the 5p state, which has a lifetime of 26 ns \cite{Gutterres:2002}. In doing so we choose a blue detuning to avoid all the lower lying hyperfine levels of the 5p state as shown in Fig. \ref{exc}. For large enough detunings $\delta$ the intermediate state can be eliminated and one ends up with an effective two level system with a total coupling strength $\Omega_R=\sqrt{\Omega_r^2 \Omega_b^2/4 \delta^2\,+\,\Delta^2}$. To obtain coherent coupling between the ground state and the Rydberg state it is necessary to achieve an effective Rabi-frequency $\Omega$ well larger than the linewidth of the Rydberg state (3.76 kHz for 43 S) or of the driving laser fields. This is easily achieved for the 5s-5p transition with only a few milliwatts of laser power but the dipole matrix element for the second step (5p-43s) is roughly $10^3$ times smaller requiring $10^6$ more laser power for a comparable coupling strength.

The typical excitation scheme excites in a first step a spin polarized sample in the $5S_{1/2}\,F=2,m_F=2$ state into the $5P_{3/2}\,F=3,m_F=3$ state and in the second step into the $nS_{1/2}\, J=1/2,m_j={1/2}$ Rydberg state. Note that the hyperfine structure of highly excited state can not be resolved and we return here to the $J,m_J$ basis. Following the standard notation \cite{Steck:2010,Sobelman:1992} we can compute the dipole matrix element for the second step by

\begin{align}
&\langle F m_f | e r_q | J^\prime m_{J^\prime}\rangle = \langle J\|er\|J^\prime\rangle \,\times\\
&\sqrt{(2F+1)(2J+1)} (-1)^{J+J^\prime-I-1+m_F+m_{J^\prime}+q}\, \times \nonumber\\
&\begin{pmatrix}
  J & I & F \\
  m_{J^\prime}+q & m_F-m_{J^\prime}-q & -m_F  \nonumber
\end{pmatrix}
\begin{pmatrix}
  J^\prime & 1 & J \\
  m_{J^\prime} & q & - m_{J^\prime}-q \nonumber
\end{pmatrix}
\end{align}

The reduced matrix element $\langle J\|er\|J^\prime\rangle$ is related to the pure radial part $\langle R_{nL}|er|R_{n^\prime L^\prime}\rangle =\int R_{nL} r R_{n^\prime L^\prime} r^2 dr$ as

\begin{align}
&\langle J\|er\|J^\prime\rangle=\langle R_{nL}|er|R_{n^\prime L^\prime}\rangle\, \times\\
&\sqrt{(2J^\prime+1)(2L+1)}(-1)^{J^\prime+L+S+1+(L^\prime-L+1)/2}\,\times\nonumber \\
&\sqrt{\frac{L_{max}}{2L+1}}
\begin{Bmatrix}
  L & L^\prime & 1 \nonumber\\
  J^\prime & J & S
\end{Bmatrix}
.
\end{align}

For the transition form the 5P$_{3/2}$ into the 43S$_{1/2}$ state we obtain the purely radial matrix element $\langle R_{5,1}|er|R_{43,0}\rangle =0.01786$ ea$_0$ leading to a total dipole matrix element of

\begin{equation}
\begin{array}[t]{l}
\langle 5P_{3/2}, F=2, m_F=2 | e r_{q=+1} | 43S_{1/2}, J^\prime=1/2, m_{J^\prime}=1/2\rangle \\ = 0.01031 ea_0
\end{array}
\end{equation}

State of the art diode laser systems based on tapered amplifiers with subsequent frequency doubling deliver about $P=300$ mW of blue light, which corresponds to an intensity of $I=9.55$ MW/m$^2$ at the center of a focused Gaussian beam with a width of $w_0=100 \mu m$. With the electric field $E=\sqrt{2 I/\varepsilon_0 c}$ we can compute the coupling strength $\Omega_b=dE/h=  11.2$ MHz. In combination with a coupling strength for the first transition of $\Omega_R=100$ MHz and a detuning $\delta=400$MHz we obtain a resonant ($\Delta=0$) two photon coupling strength $\Omega_R=1.4$MHz.

\begin{figure}
\includegraphics[width=\linewidth]{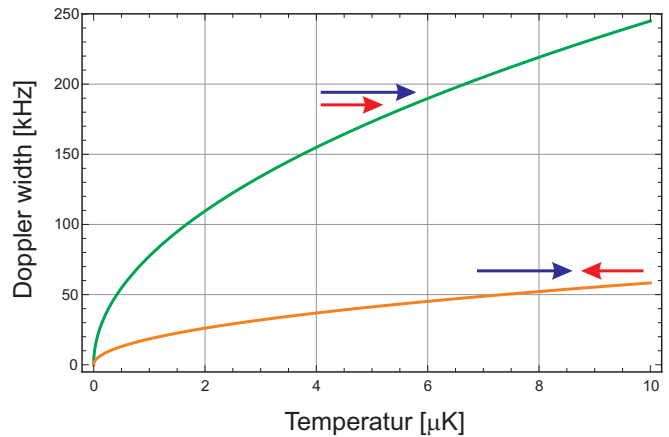}
\caption{\label{pic_doppler} The Doppler broadening depends on the directions of the incident laser beams. For parallel irradiation the Doppler shift of the two wavelengths adds up, whereas in the counter propagating case it is partially compensated. The Doppler width has to be added to the lifetime limited Lorenzian lineshape of typically narrow Rydberg lines (e.g. $ 3.8$ kHz for the 43S state) in terms of Voigt profiles. }
\end{figure}

For higher intensities one can consider dye jet lasers or dye amplifiers based on Bethune cells where the latter has been shown to produce coupling strengths in the GHz regime \cite{Huber:2011}. \\
Another important consequence of rather narrow linewidths of Rydberg states is that the Doppler broadening becomes comparable to the temperature of ultracold atomic clouds in the microkelvin regime. In Fig. \ref{pic_doppler} the Gaussian widths $\Delta\nu=\sqrt{8 k_b T \ln 2/mc^2} (\nu_b \pm \nu_r)$ of a Doppler broadened spectrum for a colinear ($\rightrightarrows$) and a counter propagating ($\rightleftarrows$) excitation are shown, which has to be convoluted with the $T=0K$ spectrum. Here $h \nu_r$ and $h \nu_b$ are the energy spacing of the lower transition and respectively the upper transition.

\subsection{\label{ChapSetup} Experimental setup for frozen Rydberg gases}

\subsubsection{\label{ChapLaser} Laser system}

\begin{figure}
\includegraphics[width=\linewidth]{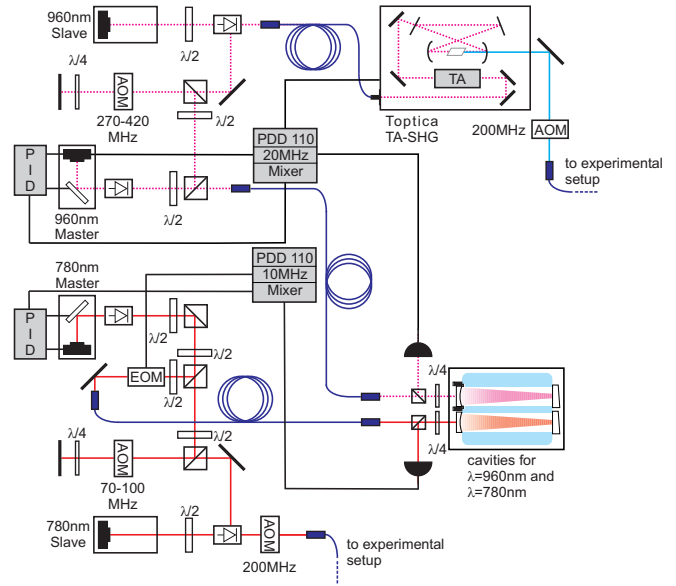}
\caption{\label{piclasersystem} Laser-system for two photon
excitation of $^{87}$Rb into Rydberg states. The red light at 780 nm
is generated by a standard external cavity diode laser system. The
production of blue light in the range of 475 nm to 483 nm is more
involved. For this purpose we use a master slave setup, where a
standard diode laser setup delivers infrared light at 960 nm which
is amplified by a tapered amplifier. A subsequent frequency doubling
cavity delivers the desired wavelength. Both lasers are stabilized (780 nm and 960 nm)
to a passively stable resonator. }
\end{figure}

\begin{figure}
\includegraphics[width=\linewidth]{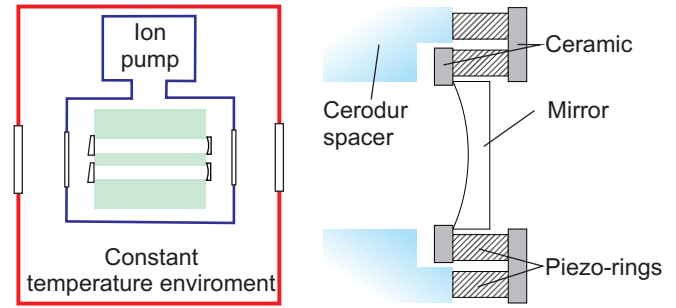}
\caption{\label{Resonator} The stability of the laser system is derived from the passive length stability of a Zerodur-glass spacer of a Fabry-Perot interferometer (FPI). To avoid thermal drifts the whole FPI is placed inside a UHV-chamber which is again place in a temperature stabilized environment. The length of the plan-concave resonator is 100mm and the finesse has been measured to $F=626$. The spacer contains in total four FPIs where two of them are additionally equipped with Piezo rings for fine tuning of the length. To minimize thermal drifts the Piezo rings are mounted in a self compensating way. }
\end{figure}

As already mentioned above we excite Rydberg states with principal quantum numbers
ranging from $n=20$ up to the ionization threshold by a two-photon excitation scheme. The lower
transition from the 5S ground state to an intermediate 5P state is accomplished with a wavelength
at 780 nm and the second step to the desired Rydberg state with roughly 480 nm.
Light at 780 nm and 960 nm is produced by standard extended cavity diode laser setups in master slave configuration. The slave lasers are seeded via acousto-optical modulators (AOM) in double-pass configuration which allows to tune the light frequency over a range of 60 MHz and 320 MHz, respectively, within a few microseconds.
The infrared light is further amplified using a tapered amplifier and then frequency doubled (TA-SHG 110, Toptica Photonics AG, Germany) to generate the needed light at 480nm. \\
Note that the experimental results on strongly interacting Rydberg gases \cite{Heidemann:2007,Raitzsch:2008,Heidemann:2008,Raitzsch:2009,Loew:2009} have been accomplished with
a precursor to our current setup based on a transfer cavity with a total linewidth of $\lesssim$1 MHz. With the new setup shown in Fig. \ref{piclasersystem}
we were able to reduce the linewidth of the combined two step excitation scheme to 60 kHz, which is an improvement of more than one order of magnitude compared to the old setup.
The key element of the new setup are two passively stable reference cavities \cite{Schulz:2009} on which we lock both lasers respectively. Therefore we use a Pound-Drever Hall-scheme \cite{Black:2001} with a fast feedback (bandwidth$\approx$10 MHz) on the current of the laser diode and a slow feedback on the grating of the extended cavity.
To maintain a preferably constant length of the plano-concave cavities, the mirrors with
a reflectivity of 0.995 were mounted on one single Cerodur spacer with a length of 100 mm which corresponds to a free spectral range of 1.50 GHz and a Finesse of 626.
To avoid thermal drifts we mounted the cavity inside a high vacuum chamber, vibrationally isolated by nitrile o-rings, at a pressure better than $10^{-8}$ mbar. Finally the whole setup is set into an actively temperature stabilized case. One concave mirror is additionally equipped with piezos to tune the length of the resonator. This setup is suitable to hit all the desired wavelengths for arbitrary Rydberg states, but may be bypassed by the usage of a more broad-band AOM-based setup, the sidebands of widely tunable EOMs, or by locking to higher TE modes of the cavities. The latter comes along with the problem that each TE mode exhibits a slightly different lineshape requiring adapted parameters for the PID locking circuit. By the special arrangement of the piezos, which exhibit a 100 times larger temperature coefficient than Cerodur, shown in Fig.~\ref{Resonator} one can compensate reasonably well for thermal drifts. An even more advanced setup would involve ultra-low expansion glasses (ULE) as a spacer to reduce the thermal drifts even further. Additionally we use very stable high voltage supplies (T1DP 005 106 EPU, iseg Spezialelektronik GmbH, Germany) with a voltage stability below $10^{-4}$. Since the piezos only approach their equilibrium length exponentially after a sudden change of the applied voltage we have to wait for a few hours to be within the stability range of our locking scheme. The combination of all these elements result in a frequency stability of  100kHz/mK, which can be further improved by the usage of ultra-low expansion glass (ULE) and leaving out the Piezo-rings.\\
During the experiment we can tune the frequency by the double pass AOMs between the master and slave lasers which results in an overall scan range of 640 MHz of the blue and 60 MHz of red light.\\

\begin{figure}
\includegraphics[width=\linewidth]{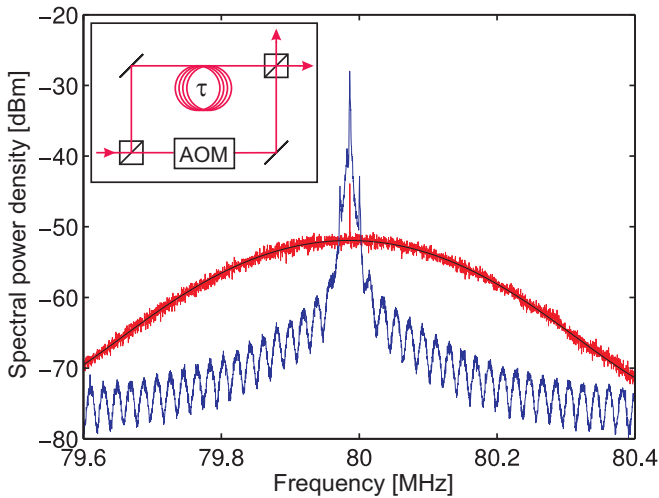}
\caption{\label{SelbstHeterodyn} Spectral power density of the beat signal from the slave laser at 780 nm with master laser locked (blue) and free running (red), measured with a delayed self-heterodyne interferometer (DSHI) as depicted in the inset. The curves are centered around the frequency of the AOM at 80 MHz. The signal of the free running  laser shows a Voigt-profile with a fitted linewidth of 210 kHz (black line) whereas the curve of the locked laser shows typical features occurring if the laser linewidth is well below the resolution limit of the DSHI (here $\approx$ 19 kHz).}
\end{figure}

The short time stability of the laser system has been studied by measuring the linewidth $\Delta f$ of the light emitted by the red and infrared slave lasers respectively using a delayed self-heterodyne interferometer (DSHI) \cite{Okoshi:1980}. The setup is depicted in the inset of Fig. \ref{SelbstHeterodyn}. In one path, the laser light is shifted in frequency using an AOM at 80 MHz; in the other arm the light is delayed using a single-mode optical fiber of 10.5 km corresponding to a delay time of $\tau=52 \text{\textmu s }$. For coherence times $1/\Delta f$ shorter than the delay time $\tau$, the two beams are completely decorrelated. The delayed light can then be treated as a second independent but otherwise identical laser. The spectrum $S_I$ of the beat signal from the DSHI is then given as the cross correlation of the original laser spectrum $S_E$ \cite{Nazarathy:1989}:
\begin{equation}
	 S_I(\omega)=\int_{-\infty}^{\infty}S_I(\omega+\omega')S_I(\omega')\,d\omega'
\end{equation}
By assuming a model for the laser spectrum $S_I$ this allows us to determine the linewidth of the laser quantitatively down to the resolution limit $1/\tau=19\,\text{kHz}$, given by the delay time. The timescale of this measurement is also determined by the delay time $\tau=52\,\text{\textmu s }$ which is adapted to the typical timescale of our experiments. For linewidths below the resolution limit, the spectrum of the beat signal shows typical features such as a well pronounced peak at the center and a modulation at the wings whose  periodicity depends on the delay time $\tau$ only. In this range only qualitatively measurements are possible unless the laser shows a pure Lorentzian spectrum \cite{Gallion:1984}. Both diode lasers in our new setup show such features. Considering the frequency doubling and the residual broadening of AOMs used for switching we can state that combined linewidth of our Rydberg excitation laser system is below 60 kHz, which agrees with the observed linewidths of experimental Rydberg spectra.  \\

In the end the light is brought to the atoms by polarization maintaining fibers and overlapped by dichroic mirrors.
At the position of the atoms the 1/e$^2$ radius of
the 780 nm light was set to 550 \textmu m and the one at 480 nm to
35 \textmu m. The maximum available laser power in the blue light (480 nm) is
about 100 mW  at the position of the atoms. In most experiments the power of the 780 nm light
is reduced well below one milliwatt to avoid excitation into the $5P_{3/2}$ state.
At a detuning of 400 MHz (see Fig. \ref{picexcite}) and a typical laser power
of e.g. 50 \textmu W the spontaneous scattering rate reduces to below one 1 kHz. The effective
two photon Rabi frequency at this setting is 250 kHz.
Another important aspect is the uniformity of the illumination of the atoms. An atomic cloud
at e.g. 3.4 \textmu K confined in our magnetic
trap at an offset field of 0.89 G, as used in \cite{Heidemann:2007}, has a Gaussian shape with a radial
width of $\sigma_\rho=8.6$ \textmu m. At this parameters
85 \% of the atoms experience at least 80 \% of the maximum
two photon Rabi frequency.

\subsubsection{\label{ChapChamber} Main vacuum chamber}

\begin{figure*}
\includegraphics[width=\linewidth]{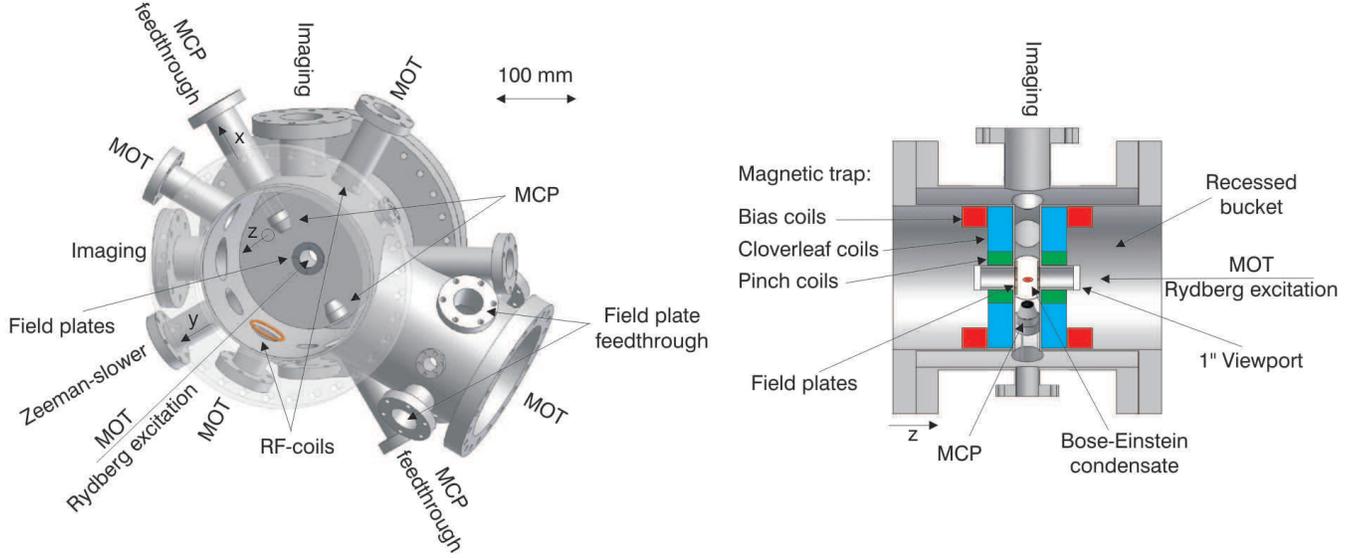}
\caption{\label{picmainchamber} Detailed view of the main chamber.
In the left part of the figure one recessed bucket is removed to
reveal the components used for electrical field manipulation and
Rydberg atom detection. The right part of the figure shows a slice
of the main chamber including both recessed buckets. Inside the
buckets, but outside the vacuum, the coils for magnetic trapping
are located. More details are given in the text.}
\end{figure*}

Our vacuum chamber for exciting and detecting Rydberg atoms in an ultracold cloud of atoms is inspired by the design of the MIT group described in \cite{Streed:2006}. The vacuum chamber consists mainly of two parts, where the main chamber is shown in Fig.~\ref{picmainchamber}, which includes now field plates to apply various electric fields and multi-channel plates for detection of charged particles. Not shown is the effusive oven assembly, which is operated at high
vacuum ($10^{-7}$ mbar) and delivers a thermal beam of gaseous Rubidium
atoms to an increasing field Zeeman slower. The precooled atoms are loaded into a magneto-optical trap in the ultra-high vacuum part of the setup ($<2\cdot 10^{-11}$ mbar). After a short molasses we transfer the atoms into a magnetic trapping potential. By forced evaporative cooling we produce for most experiments described here ultracold samples of several $10^6$ atoms at a few $\mu$K in the $F=2,\, m_F=2$ state. This is done to obtain a pure Gaussian density distribution in the harmonic trapping potential, which simplifies the subsequent analysis of our data. Typically, we work at peak densities of up to $10^{14}$ cm$^{-3}$ which is about four orders of magnitude larger than in experiments with laser cooled samples. If desired we can also extend our evaporative cooling to quantum degeneracy and produce Bose-Einstein condensates with a few $10^5$ atoms. Note that for the excitation of Rydberg states in a Bose-Einstein condensate one has to include in addition the bimodal density distribution of partially condensed clouds into the analysis as it has been done in \cite{Heidemann:2008}. \\

The main chamber depicted in Fig.~\ref{picmainchamber} has to
fulfill several boundary conditions simultaneously. First of all a
good optical access in three dimensions for laser cooling is
mandatory. Two further optical axes are added for imaging which are
equipped with larger view-ports (CF 63) to obtain a better
optical resolution. The numerical apertures in both imaging axes are
0.17 which yields an optimal resolution of 5.6 \textmu m for a
wavelength of 780 nm. Altogether eleven optical view-ports
are available which are suitable for wavelengths ranging from
300 nm to 2.5 \textmu m.  At the same time close by magnetic coils
for magnetic trapping are necessary. We use a cloverleaf style
Ioffe-Pritchard trap \cite{Pritchard:1983,Streed:2006}, which is
located inside two recessed bucket windows outside the vacuum. The
inner spacing of the two coil assemblies is 32 mm. The two pinch
coils of the cloverleaf trap produce an axial curvature of
$B^{\prime\prime}=0.56$ G/cm$^2$ per Ampere and the leaves a radial
gradient of $B^\prime=0.61$ G/cm per Ampere. At typical operation
conditions with 400 A in all coils and an offset field of 1.5 G we
obtain for $^{87}$Rb trapped in the $F=2,m_F=2$ state trapping
frequencies of 18 Hz axially and 250
Hz radially. \\

The main goal of the setup described here is the investigation of
Rydberg atoms excited from ultracold gases or Bose-Einstein condensates. For the
manipulation and detection of the highly excited atoms we included
eight field plates and two multi-channel plates (MCP) close to the
atoms inside the vacuum. The details of this add-ons will be
discussed in detail below. Because of the arrangement of the field
plates we had to relocate the radio-frequency coils used for
evaporative cooling further away from the atoms as can be seen in
Fig. \ref{picmainchamber} and Fig. \ref{picmcpplatte}. We use two
coils made of polyimide coated copper wires consisting each of two
loops. Remanent charge on the insulating coating can cause
disturbing electric fields if they are too close to the atoms, which
is also avoided by the larger distance of the coils. Nevertheless,
the coils produce at the position of the atoms an average magnetic
field of 5 mG (for frequencies from 1 to 30 MHz), when driven with 2
W at 50 Ohm without impedance matching. This is sufficient to drive the magnetic dipole transitions for
evaporative cooling. \\

A very useful addition to the setup is a helical antenna operated from outside the vacuum chamber and optimized for 6.8 GHz, which corresponds to the hyperfine splitting of $^{87}$Rb. With this antenna it is possible to adjust the density of the $F=2,\, m_F=2$ atoms by transferring a certain fraction by a Landau-Zener sweep to the magnetically untrapped $F=1\, m_F=1$ state. During this process we do not observe a severe change of the temperature, which implies also an invariant density distribution. By this we avoid the influence of all size effects during Rydberg excitation.

\subsubsection{\label{ChapDetection} Electric field manipulation and detection of Rydberg atoms}

\begin{figure}
\includegraphics[width=\linewidth]{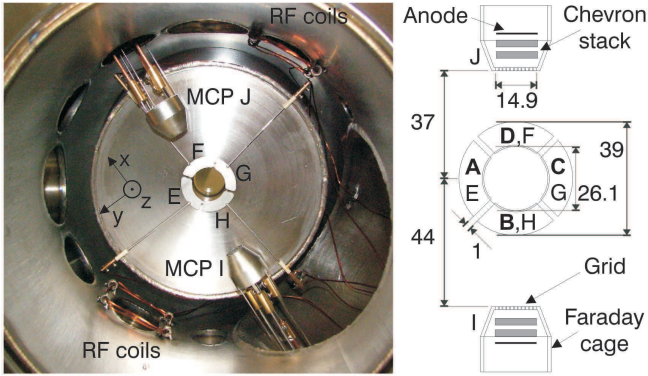}
\caption{\label{picmcpplatte} Electric field plates (A-H) and
Faraday cages ($I$ and $J$) for the multi-channel plates. Four field
plates are glued onto each of the recessed buckets, such that plate
$A$, $B$, $C$, and $D$ lie vis-a-vis to the plates $E$, $F$, $G$ and
$H$. The inner distance between the plates is 25 mm. All dimensions
given in the figure are in millimeters. The MCPs were located as
close as possible to the center of the vacuum chamber without
loosing any optical access, which resulted in the two different
distances. }
\end{figure}

The high sensitivity of Rydberg atoms to electric fields opens the
possibility to manipulate the internal states of the Rydberg atoms
by field plates \cite{Adam:2003}. To produce electric field
configurations as versatile as possible, we installed eight field
plates close to the atoms. The spatial arrangement can be seen in
Fig. \ref{picmainchamber} and Fig. \ref{picmcpplatte}. Each of these
plates can be addressed individually, which allows us to generate
various field configuration. For example for a constant electric field
of 1 V/cm pointing along the z-axis one has to charge the plates $A-D$ to -2.6 V and the plates $E-H$ to +2.6 V/cm, which produces
in addition a curvature below 0.5 V/cm$^2$ in all directions. This corresponds to an additional electric field of 50 mV/cm at a displacement of 1 mm  with respect to the geometric center. Actually we observe such an offset field around this order of magnitude \cite{Nipper:2011}, which can be explained by a spatial mismatch of the magnetic trapping potential with respect to the electric field plates. In principle it is possible to compensate for such additional fields, but it requires an elaborate calibration procedure in all three dimensions, since the direction of the perturbing field is most likely unknown. \\

The eight field plates are made of stainless steel with a thickness
of 0.5 mm. They are glued (Epotek 377, Epoxy Technology) with 1 mm thick ceramic spacers onto the
recessed buckets. The dimensions of the spacers have to be small
enough, that they are completely hidden behind the field plates from
the viewpoint of the atoms. Any insulating surface can accumulate
charge and falsify the desired field configuration. To charge the
field plates they are spot welded to stainless steel wires, which
are radially led outwards as can be seen in Fig.
\ref{picmainchamber}. At the edge of the recessed bucket the wires
are fixed in position each by a short ceramic tubing, which is also glued
to the buckets and are subsequently connected to capton insulated
copper wires. These copper wires are then finally connected to one
of the fourfold high voltage feedthroughs. To avert breakthroughs
inside the chamber induced by sharp edges we rounded off all four
edges of each plate with a radius of 1.5 mm. Finally, we etched and
electro-polished all field plates including the spot welded wires to
burnish also small spikes. The polishing was done in a acid bath
consisting of one part of 96\% sulfuric acid, two parts of 85\%
phosphoric acid and six parts of distilled water. After two minutes
at a current of 5 A about 70 \textmu m of stainless steel from the
plates was removed and they exhibited a semi gloss surface. After
installation of the field plates and evacuating the chamber we
measured no current leakage up to 3000 Volts for all plates. \\

During the experimental process it is necessary to switch the
applied voltages within short times. To do so we use bipolar high
voltage switches (HTS-6103 GSM, Behlke Electronic GmbH, Germany),
which have an intrinsic rise-time of 60 ns. The push-pull circuit of
the switch has to be adjusted to match the capacitive load of 50 pF
of each field plate as well as the 300 pF load of the high voltage coax
cable, which connects the switch to the high voltage feedthroughs.

For a high detection sensitivity of Rydberg atoms, we installed two
MCPs (Type B012VA, El-Mul Technologies Ltd, Israel) inside the
vacuum chamber. After field ionization of the Rydberg atoms, we use
one MCP to detect the ions. The second MCP is designated to detect
simultaneously the electrons. To improve the amplification even
further we use MCPs in a Chevron configuration, which consist of two
successive glass plates with a small spacing in-between. The
electron current arriving at the anode is converted by a large
resistor to a voltage and then amplified by a homebuild circuit
including a low noise operation amplifier. The whole MCP Chevron
assembly is boxed into a Faraday cage in order to shield the atoms
in the center of the chamber from the biased front side, typically
charged with -2000 V. The Faraday cage is closed at the front by a
grid with a diameter of 12 mm and a transmittance of 85\%. The
active area of the MCP front side has a diameter of 8.5 mm.

To detect the Rydberg atoms with a MCP one has first to field
ionize the excited atoms. This can be done by a large enough
electric field, which is in the case of a 43S$_{1/2}$ state about
160 V/cm. In our case we want to detect the ions and one has to
provide besides a sufficient field strength also a suitable electric
field distribution which guides the ions into the upper MCP.

Usually the magnetic fields of the trapping potential are still
switched on when the ions move towards the MCP. The combination of
electric and magnetic fields provoke a drift on the ions according
to the force $\vec{F}=q(\vec{E}+\vec{v}\times\vec{B})$. An
estimation for the given experimental situation shows that the drift
is in our case of the order of 1 mm and by this well below the
aperture of the MCP of 8.5 mm.

We calibrated the MCP ion signal by monitoring the losses in a cold
atomic cloud due to Rydberg excitation and the corresponding voltage
signal on the anode. After amplification, which is the same for the
subsequent experiments, we acquire a signal of 1 Vs per $3.65\cdot
10^{10}$ atoms. In principle one could distinguish between single
ion events, but the noise level of the signal limits our minimum
sensitivity to about one hundred ions. Since we shoot around $10^12$ ions per year into the detector we do observe aging effects of the MCP, which makes a re-calibration from time to time necessary. Due to the large distance of the MCP to the position of the atom cloud we can easily distinguish between Rb$^+$ and Rb$_2^+$ ions by their different time of flights due to their differing masses \cite{Bendkowsky:2010}.

\subsection{\label{ChapRydMT} Excitation of Rydberg states in a magnetic trap}

\begin{figure}
\includegraphics[width=\linewidth]{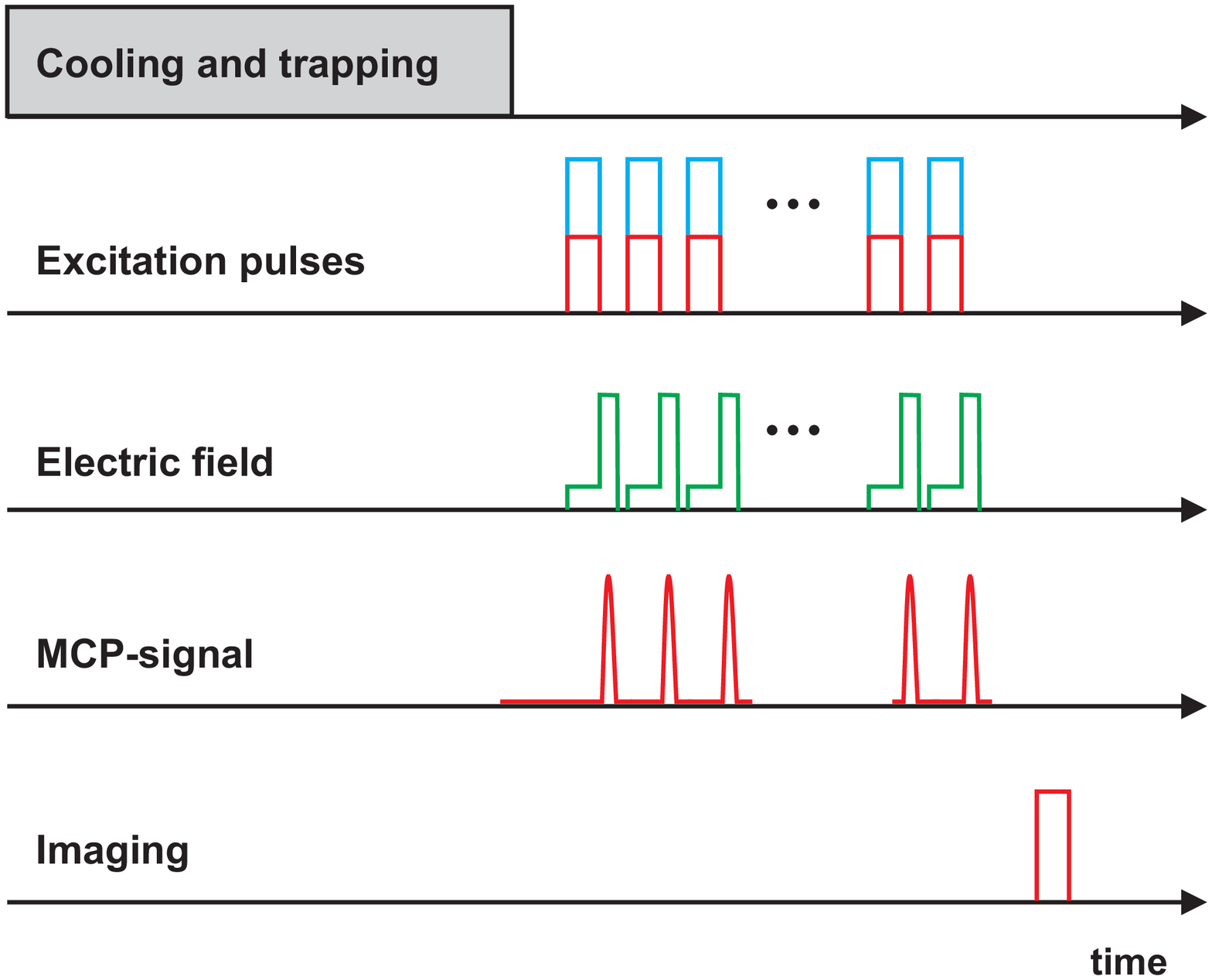}
\caption{\label{picsequence} Typical experimental sequence. In a first step we prepare within 40s a magnetically trapped cloud of ultracold atoms. During the two photon excitation we apply a small electric field to extract unwanted ions from the sample. After excitation we raise the electric field to 200 V/cm and detect the field ionized atoms with an MCP detector. The excitation and detection of Rydberg states can be repeated up to 600 times within one atomic cloud. Finally we take an absorption image of the remaining atoms.
}
\end{figure}

\begin{figure}
\includegraphics[width=\linewidth]{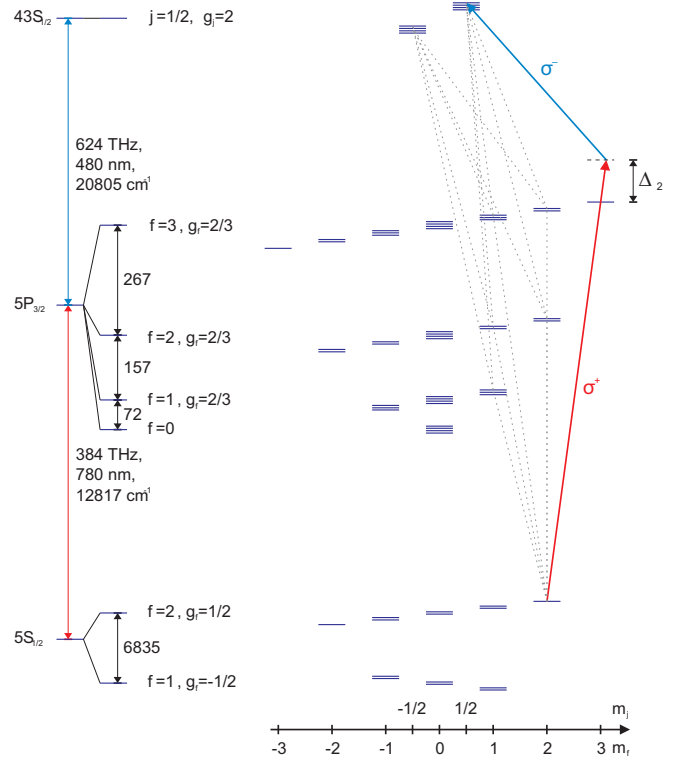}
\caption{\label{picexcite} Level scheme of the relevant levels 5S$_{1/2}$, 5P$_{1/2}$ and 43S$_{1/2}$. The left side shows a schematic of the two-photon transition including the frequency splittings (in MHz unless otherwise noted) between the hyperfine states (nuclear spin I=3/2). The right side shows the magnetic sublevels together with their respective Zeeman shift in f-basis for 5S$_{1/2}$, 5P$_{1/2}$ and in j-basis for 43S$_{1/2}$. The multiple lines denote the number of states of the respective other basis each state is composed of.
}
\end{figure}
\begin{figure}
\includegraphics[width=\linewidth]{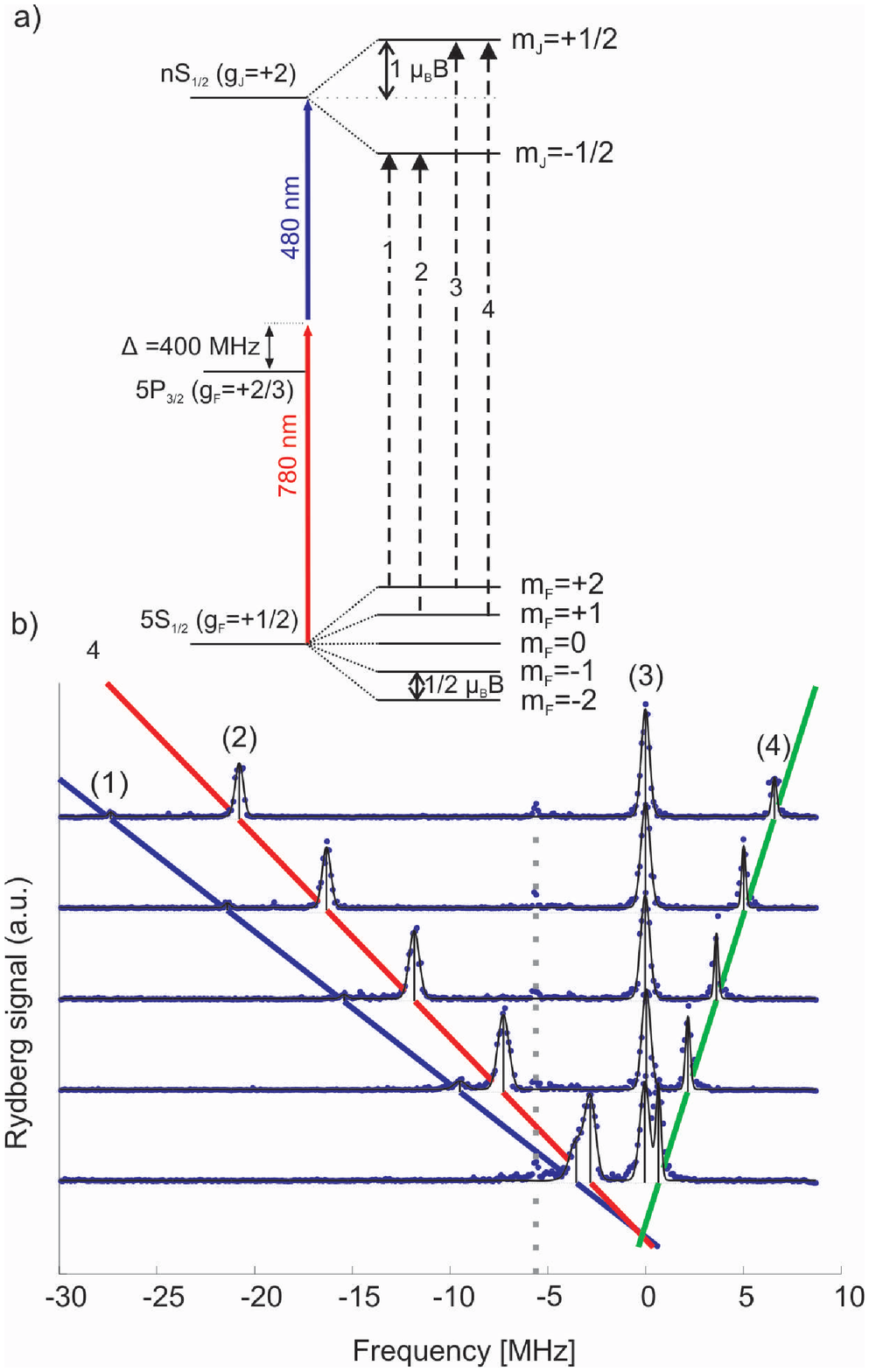}
\caption{\label{piczeeman} Excitation of Rydberg atoms into the
$43S_{1/2}$ state from a cloud of magnetically trapped atoms with a
temperature of 3 $\mu$K. a) shows the transitions from the two trapped magnetic substates of the ground state to the two magnetic substates of the Rydberg level. Transition 3 is the desired excitation path that is magnetic field independent.
b) shows spectra of Rydberg excitation at different magnetic offset fields ranging equidistant from 0.86 G to 9.34 G in a waterfall plot. Four resonances are visible whereof three are dependent on the magnetic field (see text). Solid lines show a linear fit to the peak centers, consistent with the expected Zeeman-shifts. Note that the ion detector is saturated for the strong lines in the spectra. Therefore a comparison of the transition strength is not possible. The resonance at -5.5 MHz marked by the dashed line is due to ultralong-range Rydberg molecules \cite{Bendkowsky:2009}.
}
\end{figure}

Due to the confinement of the cold atomic cloud in a magnetic trap Rydberg atoms are excited in an offset magnetic field. The experimental sequence for Rydberg excitation in a magnetic trap is shown in Fig. \ref{picsequence}. The ultracold cloud consists typically of $4\cdot 10^6$ atoms at a temperature of 3 $\mu$K confined in a cigar shaped harmonic trapping potential with an axial trapping frequency (along the z-axis) of 18 Hz and an radial trapping frequency of 310 Hz. The magnetic offset field at the center of the trapping potential was set to 1.0 Gauss during the cooling and trapping sequence. The cloud is spin polarized in the $F=2,m_F=2$ ground state with respect to the quantization axis given by the magnetic field after reloading from the MOT to the magnetic trap. Since evaporative cooling is performed via rf-coupling of neighboring $m_F$ states, a residual occupation of the weakly bound $F=2,m_F=1$ state is created in the cooling sequence.\\  For highly excited states the hyperfine coupling is significantly decreased and we remain with the coupling of the electron spin and the orbital momentum to a total momentum $J$ as shown in figure \ref{picexcite}. Atoms are excited to the Rydberg state via a two-photon excitation, detuned by about 400 MHz from the intermediate $5P_{1/2}$ state (red and blue solid arrow in figure \ref{picexcite}). The $nS_{1/2}$ Rydberg states exhibit the same magnetic moment as the ground state. Therefore the transition from the ground state to this state is insensitive to magnetic fields. Polarizations of both excitation lasers are chosen as indicated in figure \ref{picexcite} to suppress excitation via different paths (gray dashed lines) to other magnetic substates. However, the direction of the magnetic field is spatially inhomogeneous due to the geometry of the magnetic trap. This causes spatially dependent excitations into the other $m_J$ Rydberg states by admixtures of other polarizations.\\
Atoms in the $F=2,m_F=1$ ground state can also be excited to the same Rydberg states. Due to the Zeeman splitting of the ground state these transitions show different dependencies on the magnetic field. To determine the relevant transitions and explore the dependence on the magnetic offset field, Rydberg spectra at different magnetic field strength are taken as shown in Fig. \ref{piczeeman}). The magnetic offset field was ramped up shortly before the excitation and detection sequence. The spectra show four distinct resonances of which three are dependent on the magnetic field (solid lines in Fig. \ref{piczeeman}). These lines can be assigned to the expected transitions from $F=2,m_F=2$ and $F=2,m_F=1$ to $43S_{1/2},m_J=\pm 1/2$ by their magnetic field dependence. The $F=2,m_F=2$ to $43S_{1/2},m_J=+1/2$ transition, labeled by (3) in Fig. \ref{piczeeman}, is the desired excitation path.

\section{\label{ChapSiryg} Strongly interacting Rydberg gases}

  The strength of interactions in a system is usually
  characterized by the dimensionless quantity $na^3$, with $n$ being
  the density and $a$ an effective range of the interaction
  potential. Strongly interacting systems with $na^3 \gg 1$ are often
  difficult to describe theoretically. This can be understood as
  strong interactions typically lead to strong quantum correlations,
  where an effective description in terms of non-interacting
  quasiparticle excitations is no longer valid. The first steps to
  understanding the essential properties of a strongly interacting
  systems are usually based on mean-field theory, however, especially
  in low-dimensional systems this can lead to unreliable
  results. Nevertheless, it is possible to derive a mean-field
  approach to strongly interacting Rydberg gases that reproduces the
  correct critical exponents for the underlying quantum phase
  transition if the dipole blockade is explicitely taken into account
  \cite{Weimer:2008,Loew:2009}. In the case of interacting Rydberg gases
  the effective range of the van der Waals interaction between Rydberg
  states is on the order of $a\approx 5 \mu m$, which has to be
  compared with the inter-particle distance between the atoms in a
  Bose-Einstein condensate, $1/\sqrt[3]{n}\approx 100\,\mathrm{nm}$,
  leading to $na^3 = 1.25\times 10^8$, which is deep in the strongly
  interacting regime.

\subsection{\label{ChapSimpleTheo} Phenomenological Description of Strongly Interacting Rydberg Gases}

\begin{figure}
\includegraphics[width=0.5\linewidth]{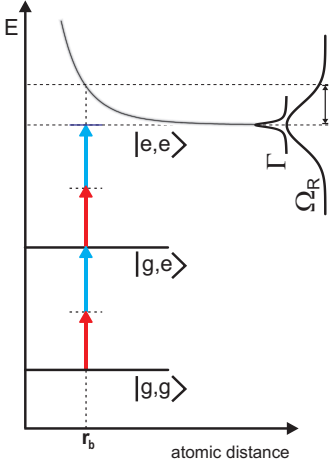}
\caption{\label{pic_blockade} Blockade of the Rydberg excitation
due to the van der Waals interaction. A
laser with a coupling strength $\Omega$ drives the transition
$|g\rangle$ to $|r\rangle$ from the ground state to the Rydberg
state. Considering two atoms, the energies
of the pair states $|g,g\rangle$ and $|g,r\rangle$ remain
almost unshifted due to the small polarisability of
the ground state atom. The transition $|g,r\rangle$
to $|r,r\rangle$ however is dependent on the interparticle
distance r because the energy level
is shifted due to van der Waals interaction
between the two Rydberg atoms. Dependent
on the saturation broadening $\Omega_R$ one can define a blockade radius $r_b$
below which the doubly excited state
is suppressed. This is true as long the coupling strength $\Omega_R$ of the excitation lasers is smaller than the linewidth $\Gamma$ of the excited state.}
\end{figure}

The interaction energy per particle in a gas of highly excited Rydberg atoms with density $n$ is given in the case of a van-der-Waals-interaction by $C_6 n^2$. For rubidium atoms in the $43S$ state $C_6$ reaches $-1.7\times 10^{19}$ a.u. = -$h\,\times$ 2441 MHz $\mu$m$^6$, which corresponds with a typical density of $n=10^{12}$ cm$^{-3}$ (mean distance $1/\sqrt[3]{n}=1\mu$m) in an ultracold atomic cloud to a pairwise interaction energy of $-h\,\times$ 2441 MHz. This energy has to be compared to the other energy scales of the system like the temperature in the $\mu$K regime, which corresponds to a kinetic energy of a few kHz, the trapping frequencies of few Hz to kHz and potentially to the chemical potential of a Bose-Einstein condensate of a few kHz. All these energies are comparable with the spectral width of the excited Rydberg states given by their lifetimes. The crucial parameter for coherent excitation dynamics is a sufficiently strong coupling strength of the exciting laser fields $\Omega_R$. The energy $\hbar\Omega_R$ determines now in combination with the interaction energy $C_6 n^2$ all bulk properties of the driven system as e.g. steady state maximum Rydberg density. By equating these two energies $C_6 n^2=\hbar \Omega_R$ for example for the 43S state and a driving field of $\Omega_R$=1MHz one can estimate a minimum distance between to Rydberg atoms of 3.67 $\mu$m ($n_r=2\times10^{10}$ cm$^{-3}$), a distance severely larger then the distance between two ground state atoms. In other words the excitation to Rydberg states is strongly blockaded and only one atom out of $n_r/n\approx$ atoms can be excited to a Rydberg state. This line of arguing is equivalent to the picture shown in Fig \ref{pic_blockade}, where the power broadening $\Omega_R$ of the exciting laser field can compensate the van der Waals repulsion up to a minimum distance. \\

\begin{figure}
\includegraphics[width=\linewidth]{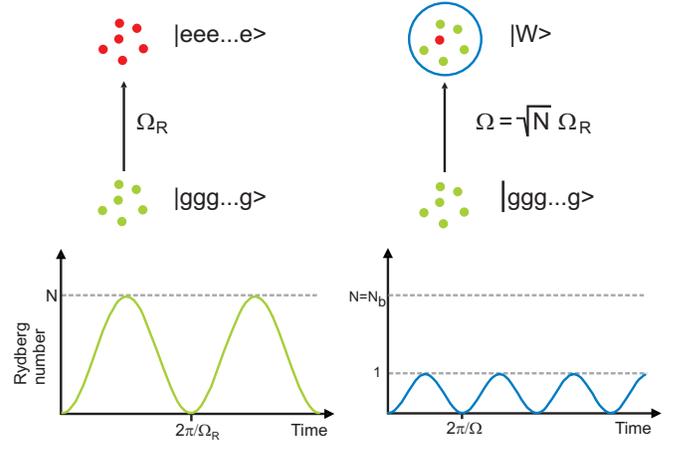}
\caption{\label{pic_collective} Resonant excitation of an ensemble of two level atoms. When exciting an ensemble of $N$ non interacting atoms with a coupling strength $\Omega$ the atoms undergo coherent Rabi-oscillations with frequency $\Omega$. In the case interactions and an ensemble size smaller than the blockade radius only one excitation is possible and the excited collective state reads $W=\frac{1}{\sqrt{N}}\sum|g_1,g_2,g_3,...,e_i,...g_N\rangle$. As a consequence collective Rabi oscillations arise with a $\sqrt{N}$ faster oscillation and by a factor $N$ reduced amplitude.}
\end{figure}

Fig. \ref{pic_blockade} gives a good illustration of the blockade mechanism during excitation of two atoms but is only correct approximately, since it neglects
the many body aspect of the excited state as described in Fig. \ref{pic_collective}. Within the blockade region one excitation is distributed over all atoms $N_b$ which results in a
collective state
$|e\rangle=|\psi_e\rangle=\frac{1}{\sqrt{N_b}}\sum_{i=1}^{N_b}|g_1,g_2,g_3,...,e_i,...g_{N_b}\rangle$, where we have dropped a spatial
dependent phase factor $\exp (-ikr)$ of the excitation light assuming a plane wave. The effective coupling $\Omega$ to the collective excited W-state shows now a $\sqrt{N_b}$ enhancement due to the collective dipole matrix element $\langle g|ed|W\rangle$ with the ground state $|g\rangle=|g_1,g_2,g_3,...g_{N_b}\rangle$. This W-state is actually the only bright state which couples to the light field, whereas the other $N_b-1$ states are dark states as can be easily seen for $N_b=2,3$ and $4$:\\

\begin{eqnarray}
N_b=2 & \nonumber\\
\mbox{Bright:} & \frac{1}{\sqrt{2}}\left(|rg\rangle+|gr\rangle\right) \nonumber \\
\mbox{Dark:}   & \frac{1}{\sqrt{2}}\left(|rg\rangle-|gr\rangle\right) \nonumber \\
\nonumber\\
N_b=3 & \nonumber\\
\mbox{Bright:}& \frac{1}{\sqrt{3}} \left(+|rgg\rangle+|grg\rangle+|ggr\rangle\right) \nonumber \\
\mbox{Dark:}  & \frac{1}{\sqrt{6}} \left(+|rgg\rangle+|grg\rangle-2|ggr\rangle\right) \nonumber \\
              & \frac{1}{\sqrt{2}} \left(+|rgg\rangle-|grg\rangle+0|ggr\rangle\right) \nonumber \\
\nonumber\\
N_b=4 \nonumber\\
\mbox{Bright:}& \frac{1}{2} \left(+|rggg\rangle+|grgg\rangle+|ggrg\rangle+|gggr\rangle\right) \nonumber \\
\mbox{Dark:}  & \frac{1}{2} \left(+|rggg\rangle-|grgg\rangle+|ggrg\rangle-|gggr\rangle\right) \nonumber \\
              & \frac{1}{2} \left(-|rggg\rangle+|grgg\rangle-|ggrg\rangle+|gggr\rangle\right) \nonumber \\
              & \frac{1}{2} \left(+|rggg\rangle+|grgg\rangle-|ggrg\rangle-|gggr\rangle\right) \nonumber \\
\end{eqnarray}

As one can easily see that in contrast to the bright W-state, the representation of the dark states is not unique. Due to the collective enhancement of the collective excitation one has to adjust the energy argument given above to a collective Rabi frequency $\Omega=\sqrt{N_b} \Omega_R$ as

\begin{equation}\label{eq.coll_block}
Z C_6 n^2 = \sqrt{N_b} \hbar \Omega_R \, .
\end{equation}

Here we introduced a coordination number which accounts for the interaction to all next and next-next neighbors for ensembles larger than the blockade radius. Assuming a crystalline ordered state of the superatoms with a hexagonal closed packing, which has the lowest energy for repulsive van der Waals interactions, the effective number of neighbors is $Z=14.4$. The number of blockaded atoms $N_b$ in each blockade sphere is given by the ratio of the density of ground state atoms divided by the density of Rydberg states $n/n_r$ and with this the density of excited Rydberg atoms and of the collective Rabi-frequency can be written as

\begin{eqnarray}
\label{eq.coll_block1}
n_r(\mathbf{r}) = \left(\frac{2 \hbar}{Z C_6} \right)^{2/5} n(\mathbf{r})^{1/5} \Omega_R^{2/5}, \\
\label{eq.coll_block2}
\Omega(\mathbf{r}) = \left(\frac{Z C_6}{2 \hbar} \right)^{1/5} n(\mathbf{r})^{2/5} \Omega_R^{4/5}.
\end{eqnarray}

\begin{figure}
\includegraphics[width=\linewidth]{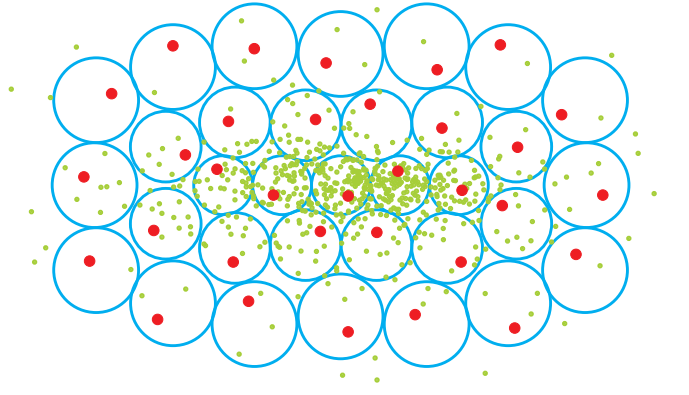}
\caption{\label{pic_inhom} Rydberg excitation of an inhomogeneous atom cloud in the superatom picture. The size $r_B$ of the blockade radius scales with the ground state density as $n^{-1/15}$ (see eq. \ref{eq.coll_block1}). When the atom number per blockade sphere approaches unity ($n_r\approx n$ then the collective description breaks down and all atoms in the dilute wings of the Gaussian density distribution have to be treated as individual atoms.}
\end{figure}

Here we have allowed for a space dependent density $n(\mathbf{r})$ which adopts, e.g., for a Gaussian distribution for a thermal sample in an harmonic trap above T$_c$ a Gaussian distribution. The Rabi frequency has been assumed to be constant over the whole ensemble which can easily be achieved by choosing sufficiently large laser beams. At large radii usually the density of the atomic cloud drops much faster than a Gaussian distribution due to a steeper potential for magnetically trapped clouds or the finite trapping region of optical dipole traps. At a certain radius the superatom model breaks down $(N_b \leq 1)$ and the evolution of the driven atoms are given by single atom Rabi frequency $\Omega_R$ which has to be included in numerical simulations by setting $\Omega=\Omega_R$ for $N_b\leq 1$. The physical situation according to the experimental approach is depicted in Fig \ref{pic_inhom} where the inhomogeneous density distribution also causes a variation of the blockade radius

We start with an inhomogeneous atomic cloud with all atoms in the ground state and then switch on suddenly the coupling light field to the Rydberg state. The evolution of the excited state fraction $f_R$, which is the total number of excited Rydberg atoms divided by the number of ground state atoms $N=\int n(\mathbf{r}) dV$, follows then

\begin{equation}\label{eq_CCRE_integral}
f_R(t)=\int \left(\frac{n_r(\mathbf{r})}{n(\mathbf{r})}\right) \sin\left(\sqrt{\frac{n_r(\mathbf{r})}{n(\mathbf{r})}} \Omega_R(\mathbf{r}) t/2\right) dV ,
\end{equation}

where we have used \ref{eq.coll_block1} and \ref{eq.coll_block2}. The integral is only within the volume where the blockade is effective. Further out one has to include the single particle dynamics separately. Note, that in this picture of noninteracting superatoms a homogeneous sample ($n=const$) $f_R(t)$ would oscillate with the collective Rabi-frequency. This is unlikely to happen for a thermal ensemble of atoms due to the random spatial normal distribution of atoms leading to a band structure of the excited states. An ordered array of atoms as e.g. given by a Mott-Insulator state will exhibit a less complex distribution of the excited states with a few discrete energies and coherent Rabi-oscillatios may be observable. The temporal evolution of $f_R(t)$ for a Gaussian density distribution results in a superposition of Rabi-oscillations with varying collective frequencies and resembles a simple saturation behavior with an initial linear increase with a rate $R$ and a exponentially approaching saturation value $f_{R,sat}$ as

\begin{equation}\label{eq_saturation}
f_R(t)=f_{R,sat} (1-e^{R t/f_{R,sat}}) .
\end{equation}

This temporal evolution agrees well with the experimental observations and also with the corresponding numerical simulations. For very short times $t<\Omega^{-1}$ the interaction energy lies below the saturation broadening and the system evolves quadratically with $f_R(t)\sim\Omega_R^2t^2$. In the case of a Gaussian distribution one can extract scaling laws for $R$ and $f_{R,sat}$ with the help of equations \ref{eq.coll_block1} and \ref{eq.coll_block2} as a function of the peak density $n_0$ and the single atom Rabi frequency $\Omega_R$

\begin{eqnarray}
\label{eq.coll_rate}
R\sim n_0^{1/5} \Omega_R^{2/5}  , \\
\label{eq.coll_sat}
f_{R,sat} \sim n_0^{-4/5} \Omega_R^{2/5} .
\end{eqnarray}

\begin{figure}
\includegraphics[width=\linewidth]{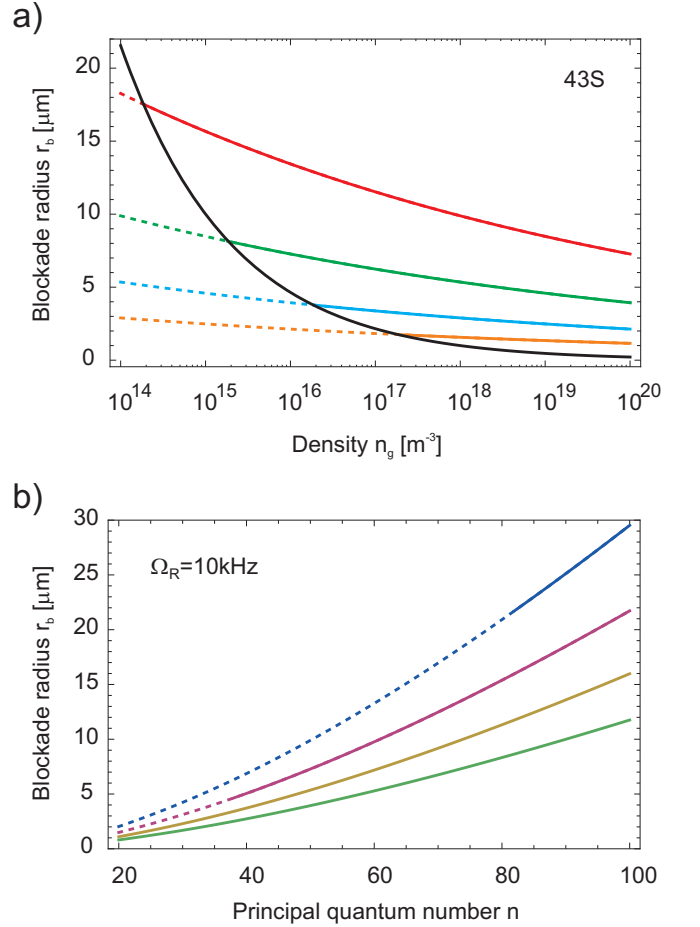}
\caption{\label{pic_blockaderadius} Fig a) shows the dependence of blockade radius for the 43S state according to equation \ref{eq.coll_block1} for a single superatom (Z=1) as a function of the ground state atom density $n_g$ and for various coupling strenghts $\Omega_R$ (red: 100Hz, green: 10kHz, blue: 1MHz, orange: 100MHz). In b) the blockade radius is plotted for S-states as a function of the principal quantum number for a fixed coupling strength of $\Omega_R=$10kHz and various densities $n_g$ (red: $10^{14}$ m$^{-3}$ , green: $10^{16}$ m$^{-3}$ ,blue: $10^{18}$ m$^{-3}$,orange: $10^{20}$ m$^{-3}$). The dashed line indicates in both figures the region, where the interparticle distance $1/\,^3\sqrt{n}$ is smaller than the blockade radius $r_b$, which is emphasized by the black line in a). }
\end{figure}

Keeping these results in mind, one notices that some care has to be
taken when investigating the strongly interacting limit given by
$na^3\gg 1$. It might be tempting to associate $n$ with the Rydberg
density $n_r$, leading to the relation $n_r r_b^3\sim 1$. Such a
reasoning, however, ignores the fact that the system is driven between
the ground state and the Rydberg state. We will show in the subsequent
theoretical analysis that the global energy scale dominating the
dynamics is $C_6n^2$ with $n$ being the ground state density. Treating
the driven system as a gas of Rydberg atoms and ignoring the ground
state atoms misses the important fact that the number of Rydberg
excitations is not a conserved quantity. This is in strong contrast to
experimental situations in which an atomic gas is brought into the
strongly interacting limit by means of a Feshbach resonance; there,
the number of atoms is indeed conserved. Nevertheless, there do exist several proposals, which try to convert
some part of these extreme interaction energies into effective interaction potential for ultracold atoms by admixing
a small fraction of Rydberg states to the ground state\cite{Honer:2010,Johnson:2010,Pupillo:2010,Henkel:2010}. To get an idea of the actual length scales of the blockade radius Fig. \ref{pic_blockaderadius} depicts its dependence for different values of the coupling field $\Omega$, the ground state density and the principal quantum number.

\subsection{\label{ChapFullTheo} Mean-Field Theory of Strongly Interacting Rydberg Gases}

The dynamics on the timescale of the experiment is well described by
the Hamiltonian in the rotating frame of the laser excitation with the
Rabi frequency $\Omega$ and the detuning $\Delta$,
\begin{equation}
  \label{eq:H} H = -\frac{\hbar\Delta}{2}\sum\limits_i \sigma_z^{(i)} +
\frac{\hbar\Omega_R}{2} \sum\limits_i \sigma_x^{(i)} + C_{p}
\sum\limits_{j<i} \frac{{P}_{ee}^{(i)}{P}_{ee}^{(j)}}{|{\bf r}_i-{\bf
    r}_j|^p},
\end{equation}
where the counter-rotating terms have been dropped
\cite{Robicheaux2005}. It is instructive to first look at the
Hamiltonian in the classical limit, i.e., $\Omega_R = 0$. For negative
detuning $\Delta$ the ground state is a paramagnet with all spins
pointing down since adding a Rydberg excitation will cost
energy. However, for positive $\Delta$ the many-body ground state will
have some Rydberg excitations present. For a fixed fraction of Rydberg
excitations $f_R$ in the system, the energy will be minimal when the
interaction energy is minimized, which is the case in a crystalline
arrangement. At $\Delta=0$ there is a continuous phase transition from
the paramagnetic to the crystalline phase \cite{Weimer:2008}. In the
strongly interacting limit it is convenient to use the interaction
energy $E_0 = C_pn^{p/d}$ as the global energy scale and express the
Rabi frequency and the detuning as dimensionless quantities, i.e.,
\begin{equation}
  \alpha = \frac{\hbar\Omega_R}{C_pn^{p/d}}\;\;\mbox{and}\;\;\eta = \frac{\hbar\Delta}{C_pn^{p/d}}.
\end{equation}

The most straightforward method to deal with a system close to a
continuous phase transition is mean-field theory. However, we cannot
simply apply conventional Landau theory, i.e., performing a Taylor
expansion of the energy in the order parameter, as there are no
correlations in Landau theory, which is incompatible with the dipole
blockade. Instead, we make an explicit ansatz for the pair-correlation
function
\begin{equation}
   g_{2}({\bf r}_{i}- {\bf
r}_{j})= \frac{\langle P_{ee}^{(i)} P_{ee}^{(j)} \rangle}{f_{R}^{2}}.
\end{equation}
In the blockaded region with the distance $|{\bf r}|$ being much
smaller than the average Rydberg spacing $a_R$ the pair correlation
function vanishes. On the other hand, at large distances $|{\bf r}|
\gg a_{R}$ the correlation disappears and consequently $g_2({\bf r}) =
1$. The transition from a strong suppression to the uncorrelated
regime is very sharp and the pair correlation function can be modeled
by a step function
\begin{equation}
g_2({\bf r}) = \theta\left(|{\bf r}|- a_R\right).
\end{equation}
Then, the mean field theory containing this pair correlation function
is obtained by replacing the microscopic interaction in the
Hamiltonian (\ref{eq:H}) by the mean interaction of the neighboring
atoms
\begin{equation}
  \label{eq:approx}
  P_{ee}^{(i)}P_{ee}^{(j)} \approx \left[P_{ee}^{(i)}f_{R}+P_{ee}^{(j)}f_R-f_R^2\right] g_2({\bf
    r}_{i}-{\bf r}_{j}),
\end{equation}
which reduces the Hamiltonian to a sum of local Hamiltonians $H_{\rm
  \scriptscriptstyle MF}^{(i)}$ at each site $i$.  In the strongly
interacting regime with $\alpha,\,\eta \ll 1$, the number of atoms in
the blockaded regime is large, i.e., $a_{R}^{d} n \gg 1$. This allows
us to replace the summation over the surrounding atoms $j$ by an
integral over space with a homogeneous atomic density $n$, i.e.,
\begin{eqnarray}
  \sum\limits_j g_2({\bf
    r}_{i}-{\bf r}_{j}) \frac{C_p}{|{\bf r}_i-{\bf r}_j|^p} &=& A_dn\int\limits_0^\infty \textup{d}^d r\, r^{d-1} g_2(r)\frac{C_p}{r^p}\nonumber\\ & =& \frac{A_d}{p-d}n{C_p}{a_R^{p-d}},
\end{eqnarray}
where $A_d$ is the surface area of the $d$-dimensional unit
sphere. Then, we obtain the Hamiltonian for the $i^{\mathrm{th}}$
atom,
\begin{align}
  \label{eq:heff} \frac{H_{\rm \scriptscriptstyle MF}^{(i)}}{E_0} &= \frac{\eta}{2} \sigma_z^{(i)} +
\frac{\alpha}{2} \sigma_x^{(i)} + \frac{A_d}{p-d} \frac{f_{R}}{n^{p/d-1} a_R^{p-d}}
P_{ee}^{(i)} \nonumber\\  & - \frac{A_d}{2(p-d)} \frac{f_{R}^2}{n^{p/d-1} a_R^{p-d}} = {\bf h}\cdot {{\boldsymbol
\sigma}}^{(i)} + h_{0}.
\end{align}
In the limit of strong blockade the correlation length $\xi$ is equal
to $a_R$ as this will minimize the interaction energy for a given
Rydberg fraction. The correlation function $g_2({\bf r})$ thus
satisfies the normalization condition
\begin{equation}
n f_R \int d{\bf r}
\left[1-g_2({\bf r})\right] = 1,
\label{eq:norm}
\end{equation}
which provides the relation
\begin{equation}
  a_{R} = (V_d f_{R} n)^{-1/d},
\end{equation}
with $V_d$ being the volume of the $d$-dimensional unit sphere. The
effective Hamiltonian is equivalent to a spin in a magnetic field
${\bf h}$, which possesses both a transverse component $h_x$ and a
longitudinal component $h_z$, and a constant energy offset $h_0$. We
can diagonalize the Hamiltonian using a spin rotation, i.e., $H_{\rm
  \scriptscriptstyle MF}^{(i)} = h \sigma_{z'}^{(i)}+h_0 $ with $h =
\sqrt{h_x^2+h_z^2}$. Here, $\sigma_{z'}^{(i)} = \cos \theta
\sigma_{z}^{(i)} + \sin \theta \sigma_{x}^{(i)}$ denotes the Pauli
matrix in the new basis while the rotation angle can be expressed in
terms of the magnetic field components as $\tan \theta =
h_{x}/h_{z}$. The ground state of the system is then given by $\langle
\sigma_{z'}\rangle = -1$, leading to the self-consistent equation
\begin{equation}
  \alpha = \frac{1}{p-d}|4 A_dV_d^{p/d-1}\,f^{p/d+1/2}-p\eta\,f^{1/2}|.
\end{equation}
This equation can be cast into the form
\begin{equation}
\alpha \sim f_R^{\delta} |B-\eta/f_R^{1/\beta}|,
\label{eq:mf}
\end{equation}
\begin{figure}[t!]
  \centering
  \includegraphics[angle=-90]{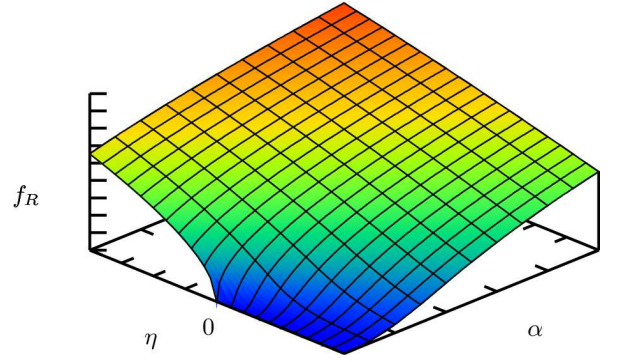}
  \caption{Excited state fraction $f_R$ of a strongly interacting
    Rydberg gas according to mean-field theory ($d=3$, $p=6$).}
  \label{fig:state}
\end{figure}
from which the critical exponents $\delta = p/d+1/2$ and $\beta = d/p$
can be determined. Solving the mean-field equation for $f_R$ leads to
the equation of state shown in Fig.~\ref{fig:state}, with the limiting
cases
\begin{eqnarray}
  f_R \sim \eta^{d/p} &\;\;\;\mathrm{for}\;\;\;& \alpha = 0\\
  f_R \sim \alpha^{2d/(2p+d)} &\;\;\;\mathrm{for}\;\;\;& \eta = 0.
  \label{eq:crit}
\end{eqnarray}
These critical exponents can also be derived using a universal scaling
function \cite{Loew:2009}.

So far, the mean-field analysis has been centered onto the ground
state properties of the system. However, in typical experiments the
Rabi frequency is not varied adiabatically such that the system
follows its many-body ground state; instead, the driving lasers are
switched to full power instantaneously, leading to a complex
relaxation dynamics towards a stationary state. The resulting critical
exponents are identical to those for the ground state. The
phenomenological superatom model can then be derived by calculating
the energy gap $h$, which for $\eta = 0$ scales like $h \sim
\sqrt{N_b}\Omega$ with $N_b$ being the number of blockaded atoms.

\subsection{\label{ChapNum} Numerical Simulations of Strongly Interacting Rydberg Gases}

To get a quantitative assessment of the validity of mean-field results
it is instructive to obtain exact numerical results on the basis of
the full spin Hamiltonian (\ref{eq:H}). For a system with $N$ atoms
the dimension of the full Hilbert space grows exponentially like
$2^N$; therefore, the exact behavior can only be calculated for a
relatively small number of atoms.  However, the strong van der Waals
repulsion suppresses the occupation probabilities of many basis
states, which allows us to significantly reduce the Hilbert space: for
each basis state we compute the van der Waals energy and remove the
state if its van der Waals energy is larger than a cutoff energy
$E_C$.  This reduction leads for $N = 100 $ to approximately $10^6$
relevant basis states compared to the $10^{30}$ basis states of the
full Hilbert space. Convergence of this method can be been checked by
increasing $E_C$.

\begin{figure}[b!]
  \centering
  \includegraphics{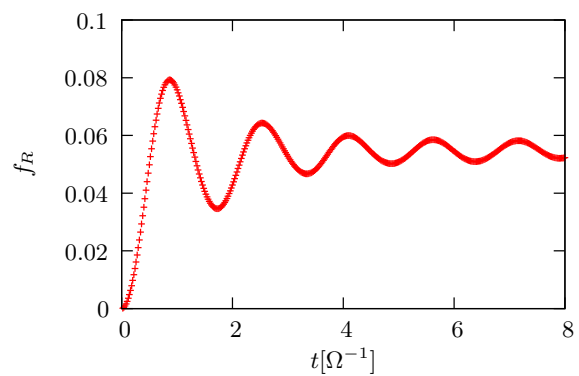}
  \caption{Relaxation dynamics of the Rydberg fraction $f_R$ ($p=6$, $d=3$, $N =
    50$, $\alpha = 1/25$, $\eta = 0$).}
  \label{fig:rk}
\end{figure}
Within mean-field theory we have already seen that the critical
properties are identical whether we study ground state properties or
the relaxation to a stationary state after a sudden increase of the
Rabi frequency. As we are also interested in dynamical properties we
perform a numerical integration of the time-dependent Schr\"odinger
equation with the initial state being the product state of all spins
pointing down. We place the atoms randomly according to a uniform
distribution into a box of unit volume having periodic boundary
conditions. Then, the resulting dynamics obtained using a fourth-order
Runge-Kutta method is shown in Fig.~\ref{fig:rk}. One can see the expected
relaxation towards a stationary state, although for $N = 50$ atoms
there are considerable finite-size effects present in the system,
especially concerning the saturation value of the Rydberg fraction
$f_R$. However, these can be reduced by averaging over many different
initial positions of the atoms according to the same distribution
function.

\begin{figure}[b]
  \includegraphics{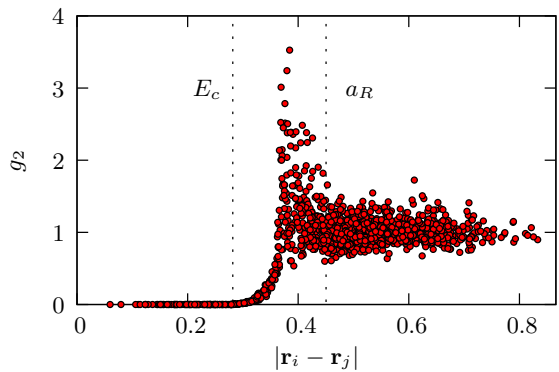}
  \caption{Pair correlation function $g_2(r)$ in the stationary state
    ($p=6$, $d=3$, $N = 50$, $\alpha = 1/25$, $\eta = 0$). The
    length scales corresponding to the cutoff energy $E_c$ and the mean
    Rydberg spacing $a_R$ are shown as vertical lines.}
  \label{fig:g2}
\end{figure}

Using our numerical simulation we can also test the validity of our
modelling of the pair correlation function as a step function. As can
be seen from Fig.~\ref{fig:g2} the assumption is qualitatively correct
although deviations can be clearly seen. There is a pronounced peak
near the correlation length $\xi$ and also the normalization
constraint of Eq.~(\ref{eq:norm}) leading to $\xi = a_R$ is also
slightly violated.

It is now possible to investigate how the scaling of the saturated
Rydberg fraction $f_{\mathrm{sat}}$ changes when the dimensionless
parameter $\alpha$ is varied. For this, the value of
$f_{\mathrm{sat}}$ needs to be averaged over many random
configurations of the atoms. Then, varying $\alpha$ by both by
changing the density $n$ and the Rabi frequency $\Omega$ leads to a
clear data collapse to single line, with the numerically obtained
critical exponents being in excellent agreement with mean-field
results \cite{Weimer:2008}.

\subsection{\label{ChapEchoSim} Simulations of Rotary Echo Techniques}

\begin{figure}[t!]
  \centering
 \includegraphics{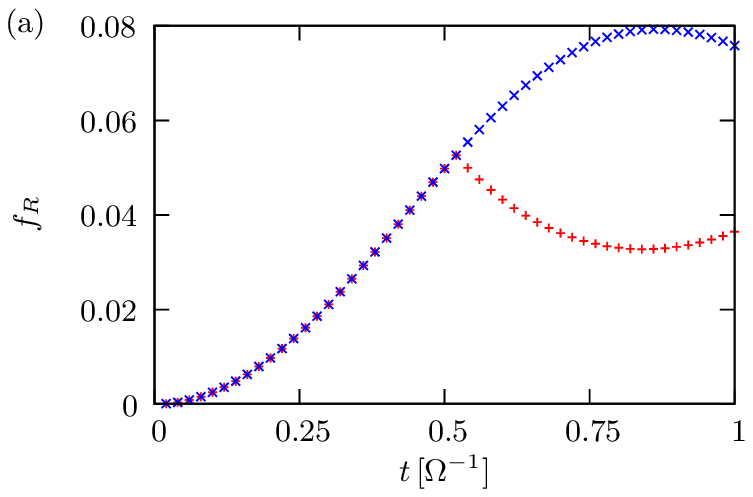}\\[0.5cm]
 \includegraphics{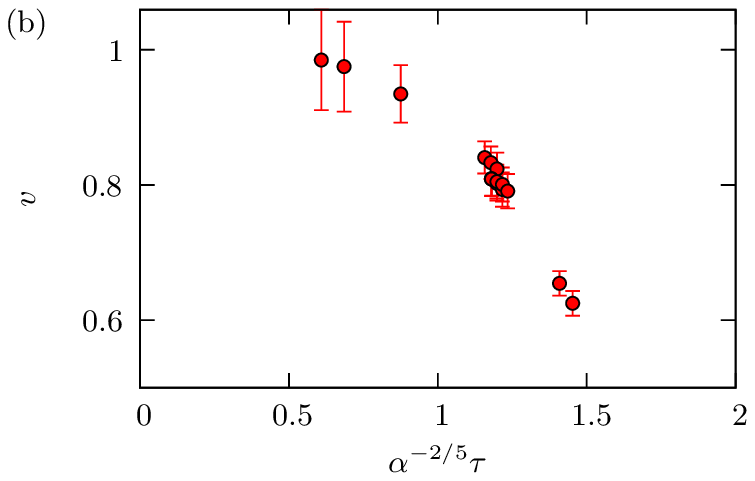}

  \caption{Numerical results for a rotary echo setup. (a)
    Time-evolution of the Rydberg fraction $f_R$ with and without
    changing sign of the Rabi frequency at $\tau_p = 0.5\,\Omega^{-1}$. (b)
    Scaling of the visibility $v$ for different times $\tau$ and
    different densities. As predicted by mean-field theory the
    visibility only depends on $\alpha^{-2/5}\tau$.}
  \label{fig:echo}
\end{figure}

A rotary echo setup can be used to investigate the coherence time of
the system subject to internal and external dephasing
processes. Examples of external dephasing are spontaneous emission and
a finite laser linewidth, whereas the dephasing of the single particle
coherence due to the van der Waals interaction acts as an internal
dephasing mechanism. In such an echo setup the Rabi frequency changes
sign at some time $\tau_p$ and the Rydberg fraction at the time $\tau
\geq \tau_p$ is measured. In absence of any internal or external
dephasing mechanisms the dynamics would be reversed, leading to
$f_R(\tau=2\tau_p) = 0$. Consequently, the Rydberg fraction serves as
a good indicator of the coherence properties of the system. In the
presence of dephasing the time $\tau$ can be chosen such that
$f_R(\tau)$ is at a minimum. Defining the Rydberg fraction for the
dynamics without flipping the Rabi frequency as $f_R'$, we can define
the visibility
\begin{equation}
  v = \frac{f_R'(\tau)-f_R(\tau)}{f_R'(\tau)+f_R(\tau)}.
\end{equation}
For simplicity, we consider a resonant driving, i.e., $\eta =
0$. Then, from mean-field theory we can expect $v$ to be a function
depending only on the intrinsic timescale
$\alpha^{-2/5}\tau$. Figure~\ref{fig:echo} shows the time evolution of
a typical echo sequence and the scaling behavior of the visibility. As
predicted by mean-field theory there is a data collapse when the
visibility is expressed as a function of the intrinsic timescale
$\alpha^{-2/5}\tau$.

The concept of the visibility can also be used to determine a
(possibly time-dependent) dephasing rate $\gamma_d$, defined by the
rate equation
\begin{equation}
  \frac{\textup{d}}{\textup{d}\tau}v(\tau) = -\gamma_d(\tau)v(\tau).
\end{equation}
Assuming $\gamma_d(\tau)$ to be a power law in $\tau$ we can directly
express the rate up to a constant factor as
\begin{equation}
  \gamma_d = -\frac{\log v}{\tau}.
\end{equation}

\section{\label{chapResults} Experimental results on Strongly Interacting Rydberg Gases}

\begin{figure}
\includegraphics[width=\linewidth]{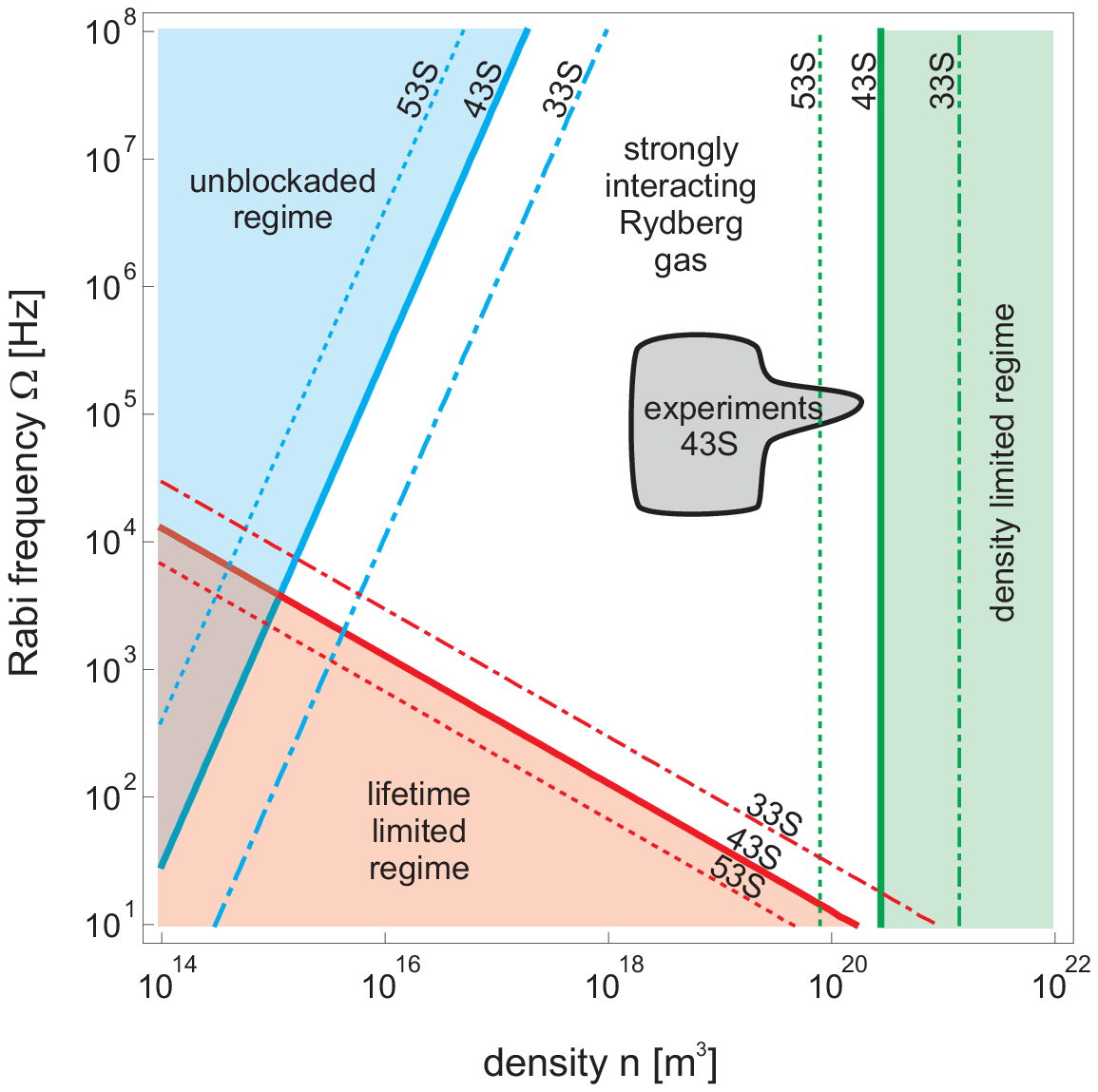}
\caption{\label{pic_Siryg} A strongly interacting Rydberg gas for e.g. the 33S, 43S and 53S state in Rubidium is limited to a certain domain in the parameter space determined by the density $n$ of the ground state atoms, the coupling strength $\Omega$ between the ground state atom and the Rydberg state, the strength of the van-der-Waals interaction $C_6$ and the excited state lifetime. The radiative lifetime limit is reached when the linewidth of the excited state $\Gamma$ is equal to the collective Rabi-frequency $\Omega_N$ and the coherent excitation dynamics gets damped out. This restriction in the time-domain can be tightened even more if a strictly frozen regime is required, where the limiting time is now given by the motion of the atoms and by this the temperature of the gas. A frozen Rydberg gas is only present as long the atoms do not move more than a wavelength of the exciting light fields during excitation. At high densities the simple description of a van der Waals interaction breaks down, due to the fact that the distance between the atoms becomes comparable to the physical size of the Rydberg atoms and the Rydberg atoms start to overlap with close by ground state atoms as it is the case for ultra-longrange Rydberg molecules \cite{Bendkowsky:2009}. Another limitation could be inelastic collisions of ground state atoms leading to three body losses \cite{Soeding:1999}. However, for the 43S state this is only relevant at densities above $10^{25}$ m$^{-3}$, which are hard to prepare in the first place. At reduced densities the system becomes weakly interacting, whenever the blockade radius $r_b$ is smaller than the distance between two atoms establishing the unblockaded regime. For some cases there will also be a limit in the regime of very high Rabi-frequencies $\Omega$ due to multi-photon ionization processes. In the regime of strongly interacting Rydberg gases represented by the white area several experiments have been conducted with various methods as described in more detail in the text. The grey shaded area represents only the parameter space of all our experiments with Rydberg atoms in the 43S state \cite{Heidemann:2007,Heidemann:2008,Raitzsch:2008,Raitzsch:2009}. }
\end{figure}

\begin{figure}
\includegraphics[width=\linewidth]{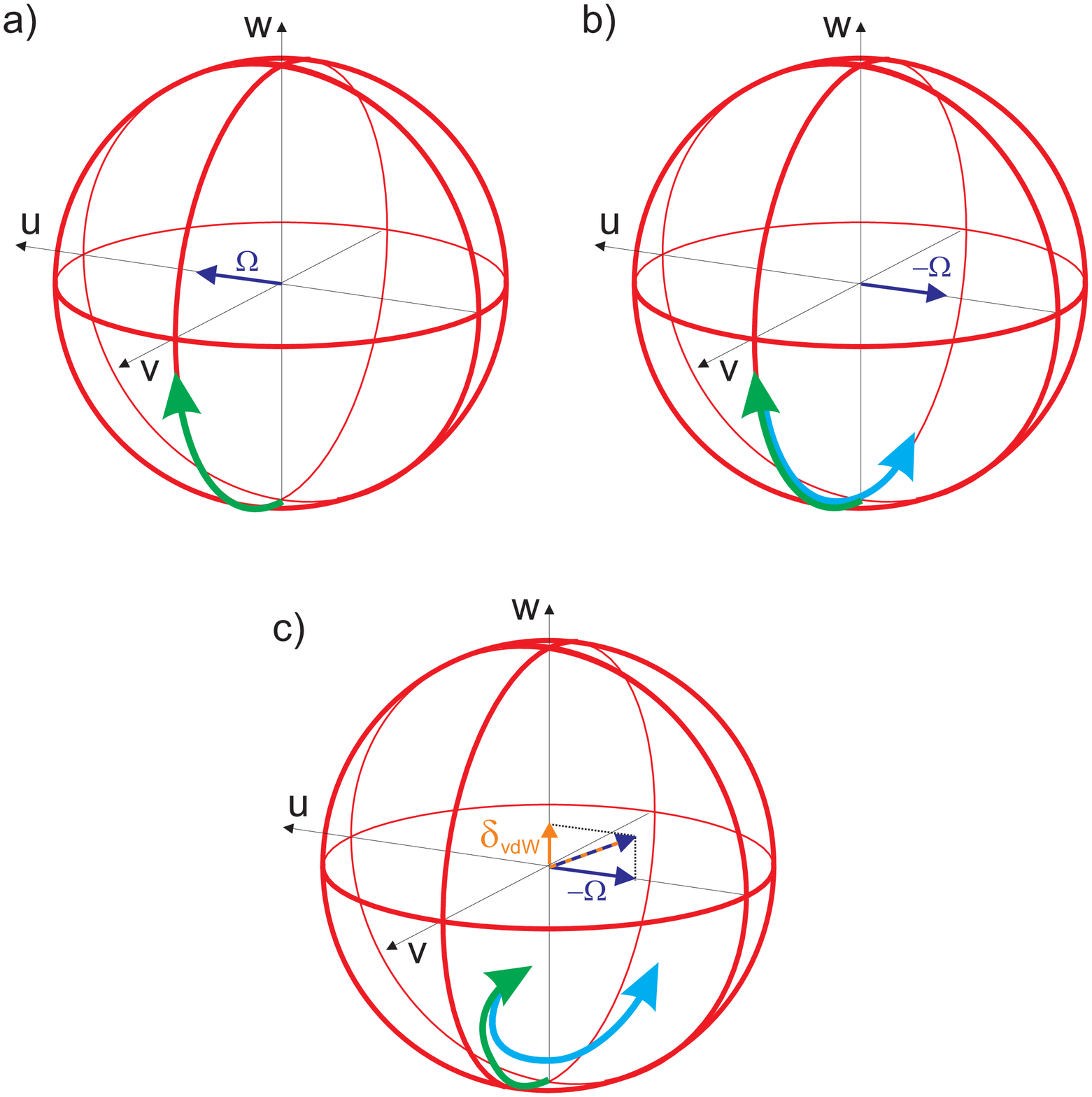}
\caption{\label{pic_blochsphere} Echo sequence depicted with Bloch-spheres. In a first step the ground state atoms are excited resonantly with a coupling strength $\Omega$ indicated by the green arrow in a). After a certain time $\tau$ the phase of the excitation field is flipped by $180^\circ$ resulting in an inversion of the excitation dynamics indicated by the blue arrow in b). In absence of interactions the population reaches after the a de-excitation time $\tau$ again the ground state. This dynamics is perturbed if with an increasing population of the excited state the van der Waals interaction between the Rydberg states induces an effective detuning $\delta_{vdW}$ as shown in panel c). The effective Rabi-frequency $\sqrt{\Omega^2+\delta_{vdW}^2}$ is now larger and points also not anymore along the $u$-axes leading to a deviation of the Bloch vector from the $vw$-plane. }
\end{figure}

Strongly interacting Rydberg are best explored in dense ultracold clouds confined either in magnetic traps or in dipole traps with various methods. But also studies on laser cooled ensembles at much lower densities give valuable insight into various interaction mechanisms and will be discussed below. Figure \ref{pic_Siryg} shows an overview over our experiments on the 43S state to date which have been conducted in the strongly interacting regime. The focus of these experiments has been lying on the coherent properties of the van-der-Waals interaction and how to control them. In our first experiment on strongly interacting Rydberg gases \cite{Heidemann:2007} we studied the excitation dynamics into Rydberg states of magnetically trapped clouds close to quantum degeneracy. In this work, as well as in all follow-up experiments, the state of choice was 43S of Rubidium, which yields a purely repulsive van der Waals potential. By this the Rydberg atoms are well protected against inelastic collisions as it happens preferential for attractive D-states \cite{Walz:2004,Robinson:2000,Li:2005,Pohl:2003,Zhang:2008}.  Nevertheless some ions are also produced during the excitation into the 43S state and it is necessary to apply some electric field ($\sim$2V/cm) to extract the ions rapidly from the atomic cloud. As a result we could observe in \cite{Heidemann:2007} a coherent evolution into the strongly blockaded regime by means of collective excitations and explained the scaling behavior within the phenomenological formalism of strongly interacting Rydberg atoms described above. Since the excitation rate and the steady state population depend on the ground state density it is possible to monitor the phase transition of thermal cloud to a Bose-Einstein condensate due to the rapid alteration of the density distribution \cite{Heidemann:2008}. Until now the coherent character of the excitation has been deduced from the scaling behavior of the excitation rate as a function of the intensity $n$ and the coupling strength $\Omega_R$. By applying more complex pulse sequences as e.g. in the case of a rotary echo it is possible to study directly coherence properties as depicted in Fig. \ref{pic_blochsphere}. This has been done for the 43S state \cite{Raitzsch:2008} and also for various D states between n=40 to n=45 in the vicinity of a F\"orster resonance \cite{Cooper:2009}.  Also electro-magnetically induced transparency has been used as a coherent spectroscopy method to investigate the influence of interactions by means of the 43S state \cite{Raitzsch:2009}. The strongly blockaded regime is accompanied by a nonuniform distribution of the excited Rydberg atoms induced by the blockade radius. A first direct observation of these correlations has been achieved by ionizing the Rydberg atoms and subsequent blow up imaging of the resulting ions by a fluorescence detector \cite{Schwarzkopf:2011}. Very recently we have used Ramsey spectroscopy to investigate F\"orster resonances in the vicinity of the 44D state.  \\

Various aspects of the interaction mechanisms among Rydberg atoms, especially F\"orster resonances, have been studied already by several groups with optically trapped atoms \cite{Reinhard:2008b,Reinhard:2008,Viteau:2011} and laser cooled atoms at somewhat smaller densities  \cite{Anderson:1998,Anderson:2002,Mudrich:2005,Vogt:2006,Vogt:2007,Reinhard:2007,Afrousheh:2004,Afrousheh:2006,Bohlouli:2007,Ryabtsev:2010,Reetz:2008}.          The sum of these experiments demonstrate the strength of resonant interactions in various approaches for different tasks, in which the studied effects become noticeable as line broadening and shifts, population transfer, and excitation blockade, which do not necessarily require throughout coherence. Nevertheless, the importance of resonant interactions lie in their tuneability, their long range character and their anisotropy \cite{Carroll:2004} and will certainly be included in various experimental settings.

\section{\label{chapConclusion} Conclusion}

This article reviews only a small fraction of modern Rydberg physics and other review articles \cite{Saffman:2010,Comparat:2010} should be consulted to obtain a more complete picture. The research field dedicated to interacting Rydberg atoms is rapidly growing and splits into self-contained research fields. A big effort is made to realize scalable quantum computers \cite{Isenhower:2010,Wilk:2010} and quantum simulators \cite{Weimer:2010,Weimer:2011} based on individually trapped single atoms, which might also be expanded to Rydberg states of ions \cite{Kaler:2011}. Actually, ensembles of strongly interacting Rydberg may also be useful as a source of cold ions since here the random potential heating is reduced by the blockade mechanism \cite{Bijnen:2011}. Although there have been quite some studies on the nature of F\"orster resonances there happens to be a lack of experiments actually using these resonances to address a specific physical problem. Possible fields of applications are quantum computing and quantum simulation. A natural research topic will be the study of energy transport in a quantum network \cite{Westermann:2006} as it happens also in many biological systems \cite{Fleming:2011}.\\
The non-linearity induced by the interactions do not only produce interesting new states of material quantum matter but will also show accordingly a  back-action on the light field. This has been observed already for laser cooled clouds with EIT-spectroscopy \cite{Pritchard:2010,Pritchard:2011} in terms of optical non-linearities. There exist various proposals to use the blockade effect as an optical non-linearity on the single photon level to generate non-classical states of light fields as e.g. Fock states \cite{Pederson:2009,Olmos:2010,Laycock:2011,Pohl:2010} and cat states\cite{Honer:2011}.\\

\section{\label{chapGlossary} Glossary}

Here we list the nomenclature of some important quantities to avoid misunderstanding due to a varying notations in other publications.\\

\begin{tabular}{llp{7cm}}
$n$ &=& Density of ground state atoms \\
$n_r$ &=& Density of Rydberg atoms \\
$\Omega_r$ &=& Coupling strength of the first excitation step (here the 5S-5P transition in Rubidium at 780 nm) \\
$\Omega_b$ &=& Coupling strength of the second excitation step (here the 5P-nS,nD transition in Rubidium at 480 nm) \\
$\Omega_R$ &=& Effective Rabi-frequency for the two photon excitation \\
$\Omega$ &=& Collective Rabi-frequency \\
$r_b$ &=& Blockade radius \\
$\alpha$ &=& Rescaled coupling strength $\Omega_R$\\
$\delta$ &=& Detuning with respect to the lower transition, here 5S-5P \\
$\Delta$ &=& Detuning with respect to the two photon transition, here 5S-nS \\
$\eta$ &=& Rescaled detuning $\Delta$ \\
\end{tabular}

\section{\label{chapAckno} Acknowledgment}

The authors thank Helmar Bender, Vera Bendkowsky, Axel Grabowski, Rolf Heidemann, Peter Kollmann, Eva Kuhnle, Ludmila Kukota, Ulrich Krohn, Johannes Nold and Alban Urvoy for their experimental and theoretical contributions. Funding for the ultracold Rubidium Rydberg setup has been provided by the DFG (Deutsche Forschungsgemeinschaft) and the European Commission. H.W.~acknowledges support by the National Science Foundation through a grant for the Institute for Theoretical Atomic, Molecular and Optical Physics at Harvard University and Smithsonian Astrophysical Observatory and by a fellowship within the Postdoc Program of the German Academic Exchange Service (DAAD).

\bibliography{litneu}

\end{document}